\documentstyle[aps,preprint,prb]{revtex}

\input{epsf}

\begin{document}

\draft
\preprint{G\"oteborg-ITP-96-10}

\title{\bf A class of ansatz wave functions for 1D spin systems and their
relation to DMRG} 

\author{ Stefan Rommer and Stellan \"{O}stlund }

\address{\normalsize\em Institute of Theoretical Physics\\ 
Chalmers University of Technology\\
S-41296 G\"{o}teborg, Sweden}

\date{\today}

\maketitle

\begin{abstract}
We investigate the density matrix renormalization group (DMRG) discovered by
White and show that in the case where the renormalization eventually
converges to a fixed point the DMRG ground state can be simply written as a
``matrix product'' form. This ground state can also be rederived through a
simple variational ansatz making no reference to the DMRG construction.  We
also show how to construct the ``matrix product'' states and how to
calculate their properties, including the excitation spectrum. This paper
provides details of many results announced in an earlier letter.
\end{abstract}


\narrowtext

\section{Introduction}

After Wilson's development of the renormalization group (RG) to solve the
Kondo problem \cite{Wilson75ren} it was believed that RG could be used for
other problems as well. Kadanoff's blocking technique combined with Wilson's
RG idea was applied to problems like quantum lattice systems such as the
Hubbard and Heisenberg models but progress turned out to be surprisingly
difficult. However, in 1992 White developed the density matrix
renormalization group (DMRG) \cite{White92den,White93den} method which since
then has had a spectacular success in calculating ground state energies and
other static properties of many 1D quantum systems.  In this paper we
explore the nature and underlying principles of the DMRG to find out why the
results of DMRG calculations are so remarkable accurate.  A summary of this
work has been presented in an earlier paper \cite{Ostlund95the}, and the
present paper provides a complete discussion and derivations of the results.
For background information on the DMRG there are excellent articles by White
\cite{White92den,White93den}.

In Sec.~\ref{sec:dmrg} we give a very brief summary of the DMRG.  In
Sec.~\ref{sec:mprod} we show that if the DMRG algorithm converges to a fixed
point, the DMRG ground state leads to a special
ansatz form for the wave function, demonstrating the equivalence of the DMRG
to a variational calculation. To make things more concrete we apply our
ideas to the antiferromagnetic Heisenberg spin-1 chain with quadratic and
biquadratic interactions, defined by
\begin{equation}
H = \sum_{i=1}^{n} {\bf S}_i \cdot {\bf S}_{i+1} - \beta ( {\bf S}_i
\cdot {\bf S}_{i+1})^2. \label{eq:heis}
\end{equation}
In Sec.~\ref{sec:q} we define a set of variational states using the special
ansatz form of Sec.~\ref{sec:mprod}.  In Sec.~\ref{sec:bloch} we extend the
ansatz to include a set of Bloch states that describe elementary excitations
in both finite and infinite systems.  These calculations are rather lengthy
and the details can be found in appendices.  Sec.~\ref{sec:results} contains
some numerical results for the spin-1 chain to compare our variational
ansatz to more involved calculations.  

We would like to mention that all
numerical work described here was programmed with Mathematica on an ordinary
desktop workstation. Each calculation we describe here take anywhere from a
few seconds to a few minutes.

\section{Density Matrix Renormalization Group (DMRG)}
\label{sec:dmrg}

Since the DMRG was discovered by S.R.\ White \cite{White92den} in 1992 it
has had great success in describing 1D interacting quantum systems
\cite{White93den,White93num,Sorensen93lar,Sorensen94equ}.  Ground state and
excited state properties have been calculated to high accuracy with modest
computational effort. With hindsight, it will be seen that the ideas of this
paper do not logically depend on the DMRG, but they were inspired by the
DMRG and we will therefore begin this section by summarizing some aspects of
the DMRG.

In a renormalization scheme like the DMRG one typically starts with a very
short 1D chain and then lets the length increase by iteratively adding a
single site. After each new site, an approximate Hamiltonian is
constructed. This is done by keeping only a small subspace of the Hilbert
space to keep the Hilbert space at a manageable size as one lets the chain
grow.  The central idea in DMRG is to keep the ``most probable'' states when
truncating the basis in contrast to the usual old-fashioned real space RG
methods (see e.g.\ Ref.~\onlinecite{White93den} and references therein)
where the lowest energy eigenstates are kept. The way to achieve this is to
split a complete system (``universe'') into two parts, a ``subsystem'' and
an ``environment'' and then to construct the reduced density matrix for the
``subsystem'' as part of the ``universe''.  The state of the ``subsystem''
is then given by a linear combination of the eigenstates of the density
matrix with weights given by the eigenvalues.

The renormalization starts with a short 1D lattice with just a few
sites. Label this system $H_B$. A renormalization step of the DMRG can be
described by the following algorithm:
\begin{enumerate}
\item Construct the Hamiltonian for the ``universe'', $H_S = H_B + H_1 +
H_1^R + H_B^R$, where $H_B$ comes from the previous iteration and $H_1$ is a
new site added. The superscript $R$ denotes a second block that is reflected
before joined to the other parts. The block $H_B + H_1$ now is our
``subsystem'' and $H_1^R + H_B^R$ our ``environment''. The Hamiltonian
matrix $H_S$ for the ``universe'' is constructed with tensor products
involving the intrablock parts $H_B$ and $H_1$ and the interactions between
the blocks.
\item Diagonalize $H_S$ to obtain the ground state $| \Psi \rangle $ of the
``universe''. This state is called the target state.
\item Construct the reduced density matrix $\rho_{i,i^\prime} = \sum_j
\Psi_{i,j} \Psi^*_{i^\prime, j} $, where $| \Psi \rangle = \sum_{i,j}
\Psi_{i,j} |i \rangle \otimes |j \rangle $ and $|i \rangle , |j \rangle $
are basis states of the ``subsystem'' and the ``environment''
respectively. The eigenstates of $\rho$ with the highest eigenvalues
correspond to the most probable states of the ``subsystem'' when the
``universe'' is in the state $|\Psi \rangle $.
\item Now choose the $m$ states of the diagonalized density matrix with
highest eigenvalues to form a new reduced basis for the block $H_B + H_1$
. Project the Hamiltonian and other operators onto this basis by
$H_{B^\prime} = A ( H_{B+1} ) A^\dagger$, where $A$ is the projection
operator one constructs from the kept eigenstates of the density matrix from
step 3 and $H_{B+1}$ is the Hamiltonian matrix for the ``subsystem''. If the
single site added has $m_s$ states in its basis, $A$ is represented by a $m
\times (m*m_s)$ matrix and $H_{B+1}$ by a $(m*m_s) \times (m*m_s)$ matrix.
\item Rename $H_{B^\prime}$ to $H_B$.
\end{enumerate}
This completes an iteration.

\section{The matrix product state} 
\label{sec:mprod}
To begin the renormalization procedure one starts with a block consisting of
a short lattice whose basis states can be calculated exactly. When the
renormalization proceeds and the chain described by the block gets longer we
don't use the full set of basis states for describing the block but have to
discard some part of the Hilbert space in each renormalization step.

Assume we have a block that represents a chain with $n-1$ sites. Let $m_s$
be the number of possible states of a singe lattice site. If we would treat
this system exactly there would be $m_s^{n-1}$ states in the Hilbert space
basis for this system. In the case of a spin 1 chain, we could label the
site with the $z$-component of the single spin-1, so that $m_s=3$. The
number of states in this complete basis rapidly becomes too large to handle
when $n$ is increased. Assume therefore that an approximation is made and
our chain is represented by a smaller set of states labeled by $\{ | \beta
\rangle_{n-1} \}$. This set of states has been chosen by the previous
iterations of the renormalization with the aim to describe the low energy
physics. Assume there are $m$ states in this basis, where $m \leq
m_s^{n-1}$. If this is the first iteration, $\{ | \beta \rangle_{n-1} \}$ is
the complete basis.

\begin{figure}
\centerline{
	\epsfxsize=7cm
	\epsfbox{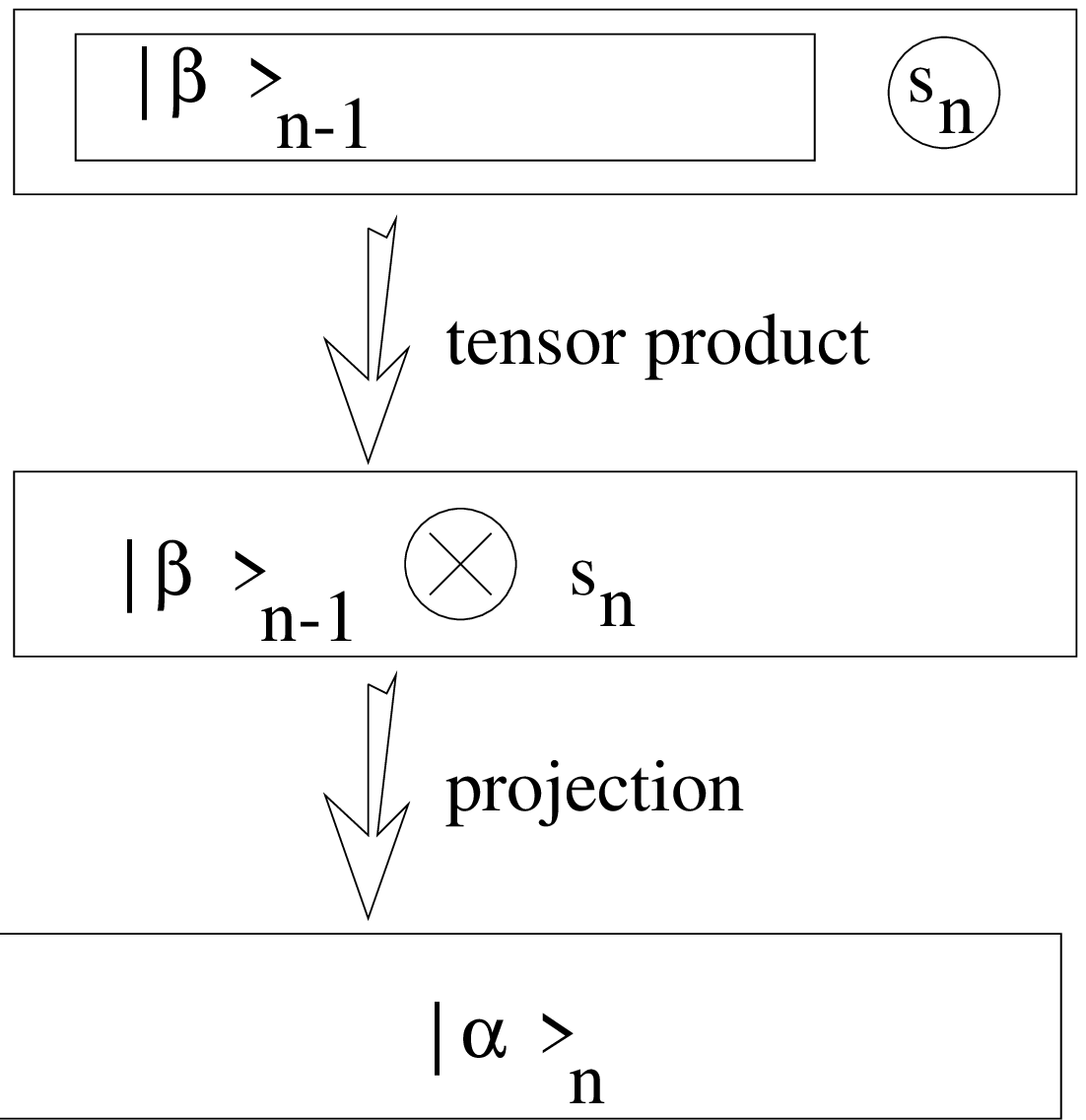}
}
\caption{A renormalization step. First the old block representing $n-1$
	sites is joined to a single site. Tensor products then ``glues'' the
	different parts together. Finally there is a projection to a new
	block representing $n$ sites.}
\label{fig:projection}
\end{figure}

We now add a single site, labeled by $s_n$, the $z$-component of spin, to
the left hand side of our block resulting in a new block with $n$ sites and
$m_s \times m$ states in its basis. The basis states are now generated by
the product representation $\{ |s_n \rangle \otimes | \beta \rangle_{n-1}
\}$. We now use a projection operator $A_n$ to generate a new truncated
basis with typically $m$ states that represent the ``important'' states of
the longer block. This whole process is written (see also
Fig.~\ref{fig:projection})
\begin{equation}
| \alpha \rangle_n = \sum_{\beta, s_n} A_n^{\alpha, (\beta, s_n)} |s_n
\rangle \otimes | \beta \rangle_{n-1} , \label{eq:renstep}
\end{equation}
where we have indexed $A$ by the chain length $n$ and its matrix indices
$\alpha$ and $(\beta, s_n)$. Note that $(\beta, s_n)$ is thought of as a
single index labeling a tensor product of the states $|s_n \rangle$ and
$|\beta \rangle_{n-1}$.

In the DMRG, a specific algorithm is used to calculate $A$, but this is not
important in the present discussion.

We now make two crucial observations. 
\begin{enumerate}
\item First we perform a simple change in notation: $A_n^{\alpha,
\beta}[s_n] \equiv A_n^{\alpha, (\beta, s_n)}$, thus writing the $m \times
(m_s * m)$ matrix as a set of $m_s$ $m \times m$ matrices.
\item Second, we assume that the recursion leads to a fixed point for the
projection operator so that we can write $A_n[s] \rightarrow A[s]$, as $n
\rightarrow \infty$.
\end{enumerate}
By recursively applying the renormalization step in Eq.~(\ref{eq:renstep})
we now find that
\begin{equation}
| \alpha \rangle_n = \sum_{s_n \ldots s_1, \beta } (A[s_n] A[s_{n-1}] 
\ldots A[s_1] )^{\alpha, \beta} | s_n s_{n-1} \ldots s_1 \rangle \otimes |
\beta \rangle_0 , \label{eq:rec} 
\end{equation}
where $| \beta \rangle_0 $ represents some state far away from $n$. We thus
see that the renormalization procedure results in a wave function that can
be written in a matrix product form. Eq.~(\ref{eq:rec}) now suggests a
natural form for the wavefunction with the following ansatz.

For every $m \times m$ matrix $Q$, we define the (unnormalized) state
$|Q)_n$ 
\begin{equation}
| Q )_n \equiv \sum_{ \{ s \} } { \, \mbox{tr} \left( \;   Q A[s_n] \ldots
A[s_1]  \; \right) } | s_n \ldots 
s_1 \rangle .
\label{eq:qstate}
\end{equation}
Thus $|Q)_n$ can be viewed as a state that is uniform in the bulk, but with
a linear combination of boundary conditions defined by $| \alpha \rangle_n $
on the left and $ | \beta \rangle_0 $ on the right \cite{Qin94edg}. The
special case of $ Q= \openone $, the identity matrix, leads to a state with
periodic boundary conditions. This $Q = \openone$ state we will later on use
as our trial ground state.

If we now demand that the projection of Eq.~(\ref{eq:renstep}) preserves
orthonormal bases, $\langle \alpha | {{\alpha}^{\prime}} \rangle =
\delta_{\alpha, {{\alpha}^{\prime}} }$, we can use the recursion formula
Eq.~(\ref{eq:renstep}) and the orthogonality of the local spin states and
previous block states to find
\begin{eqnarray}
\delta_{\alpha, {{\alpha}^{\prime}} } & = & \sum_{ \beta,
{{\beta}^{\prime}}, s, {{s}^{\prime}} } (A^{{{\alpha}^{\prime}},
{{\beta}^{\prime}}} [{{s}^{\prime}}] )^* A^{\alpha, \beta}[s] \; \langle
{{s}^{\prime}} | s  \rangle \langle {{\beta}^{\prime}} | \beta \rangle
\nonumber \\  
& = & \sum_s (A[s] A^\dagger [s])^{\alpha, {{\alpha}^{\prime}} }
. \label{eq:ortho} 
\end{eqnarray}
Hence in matrix form we have $\sum_s A[s] A^\dagger [s] = \openone $. This
constraint will be used later to reduce the number of free parameters in
$A$.

\label{sec:constrofa}

We now analyze the projection matrix $A$.  The Hamiltonian of
Eq.~(\ref{eq:heis}) is spin rotationally invariant since it commutes with
all three components of the total spin ${\bf S}_{tot} = \sum_i {\bf
S}_i$. In order that the projection in each step preserves this symmetry,
our basis states of a block must form a representation of total spin. Since
we keep basis states with many different values of total spin as well as
many states with the same total spin in each iteration, all the basis states
together must form a sum of irreducible representations of total spin.
Adding a spin one does not mix even or half-odd spin representations, thus
the basis states must form a sum of either all half-odd or all integer spin
representations. Most naturally for the spin-1 chain one would work with
integer spin representations, but by placing a single spin-1/2 on the right
hand side of the entire chain one could use half integer spin
representations to represent the blocks instead. This is consistent with the
existence of a spin-1/2 edge state \cite{Qin94edg,AKLT88val} and we have
found that working with half-odd integer representations give far better
numerical results.

\begin{figure}
\centerline{
	\epsfxsize=\textwidth
	\epsfbox{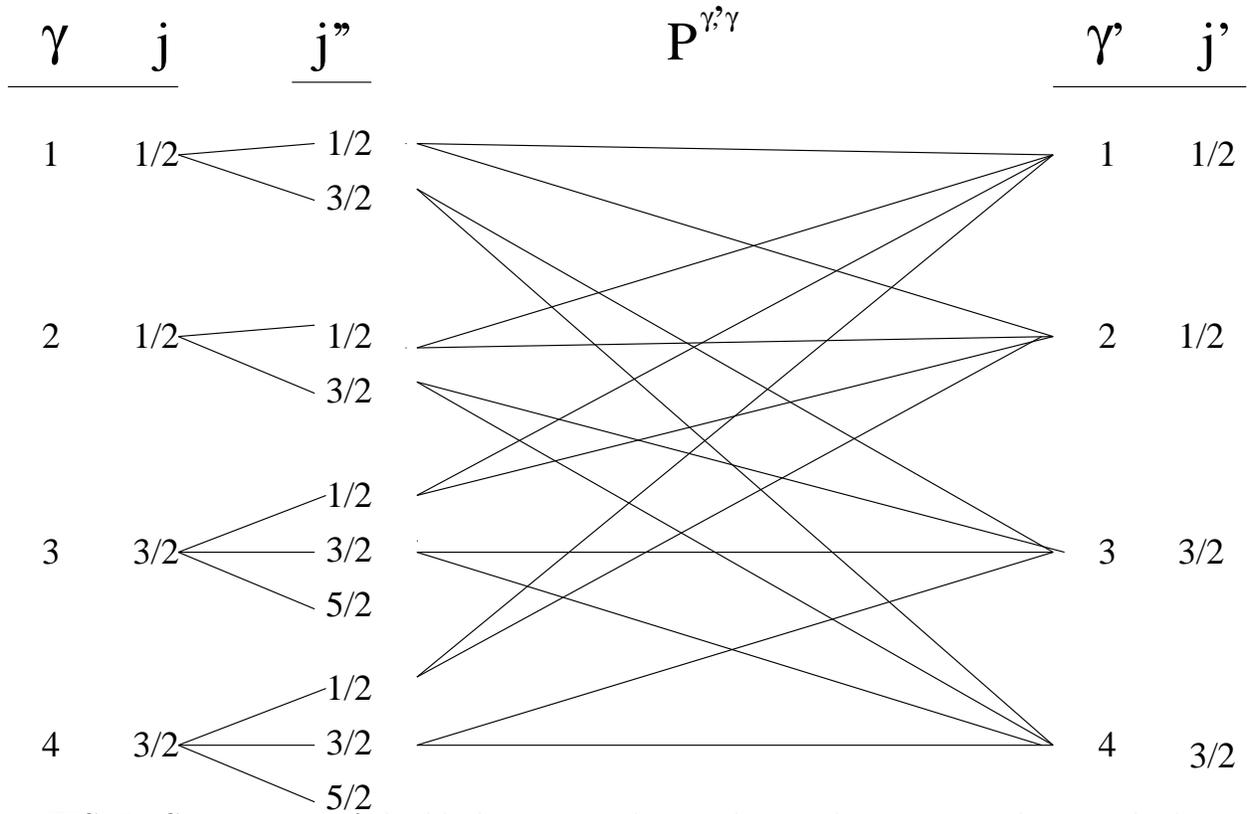}
}
\caption{Construction of the block states is shown when twelve states are
	kept in the basis. Old representations are on the left and new
	representations on the right. Each line represents a nonzero
	projection $P^{{{\gamma}^{\prime}},\gamma}$ of basis
	representations. The $z$-component of total spin is not explicitly
	written out.}
\label{fig:connect}
\end{figure}

We now discuss how the projection operator $A$ can be constructed. In our
numerical work, we have kept 12 basis states in each iteration and we have
used the half-odd spin representations. By doing a DMRG calculation on the
spin-1 chain we have found that when approximately 12 states are kept the
blocks are represented by a sum of two spin-1/2 and two spin-3/2 irreducible
representations. Since there are two sets of each representation, we have to
introduce a new label, $\gamma$, to distinguish them. The ``old''
representations representing the old block we have uniquely labeled by
ordinal number $\gamma$, with the corresponding total spin $j$. (See
Fig.~\ref{fig:connect}.) Implicit in the labeling of the states is the
$z$-component of total spin $m$, which is not shown in
Fig.~\ref{fig:connect}. These are thus the twelve ``old'' states $|\gamma, m
\rangle $, that fall into the four different irreducible representations of
total spin.

After adding a single site and then truncating the Hilbert space we get the
``new'' basis states similarly labeled by ${{\gamma}^{\prime}}$ and their
corresponding total spin ${{j}^{\prime}}$, to the right in
Fig.~\ref{fig:connect}. These will thus be the twelve basis states
$|{{\gamma}^{\prime}}, {{m}^{\prime}} \rangle$ that represent the new block
with one more site.

Let us now examine what happens in our example when going from the old
$\gamma$ to the new ${{\gamma}^{\prime}}$.  When adding a single spin-1 to
the old block of twelve states we get 36 ``intermediate'' states in the
product representation of the old block states with a spin 1. These states
fall into 10 irreducible representations labeled in Fig.~\ref{fig:connect}
by their total spin $j^{\prime \prime}$. We then project from these 10 reps
back down to the four reps that we have chosen to keep. This projection must
preserve the total spin symmetry, i.e.\ it can't mix different $j^{\prime
\prime}$ and cannot depend on total $m^{\prime \prime}$. The only nonzero
projection terms $P^{{{\gamma}^{\prime}},\gamma}$ are indicated by lines in
the figure. Since there is exactly one ``intermediate'' spin-1/2 and one
spin-3/2 for each of the four ``old'' representations $\gamma$ there is one
projection term from each of the ``old'' $\gamma$ to each of the ``new''
${{\gamma}^{\prime}}$. There are thus 16 nonzero projection terms which are
in fact not independent, but are related by the requirement that the new
states are orthonormal.

Expressing all this mathematically, we let, as in the figure, $\gamma$
uniquely label a rep of total spin of a block and $j(\gamma)$ denote the
value of total spin of that rep. Each state is thus labeled by $|\gamma,m
\rangle$ where $m$ is the $z$-component of total spin. The single spin to be
added is labeled by $| s \rangle$, where $s$ is the $z$-component of the
spin-1. The new states are thus given by
\begin{equation}
| {{\gamma}^{\prime}}, {{m}^{\prime}} \rangle = \sum_{\gamma}
P^{{{\gamma}^{\prime}},\gamma} | \gamma, 
j({{\gamma}^{\prime}}), {{m}^{\prime}} \rangle , \label{eq:proj}
\end{equation}
where $| \gamma, j({{\gamma}^{\prime}}), {{m}^{\prime}} \rangle $ denotes
the 36 intermediate states formed by $| s \rangle \otimes | \gamma, m
\rangle$ written in the total spin basis. Since we demand that the
projection preserves total $j$ and $m$, these states can be explicitly
constructed using the Clebsch-Gordan coefficients on the form $ \langle
(j_1, m_1) (j_2, m_2) | j, m \rangle$ as
\[
| {{\gamma}^{\prime}}, j({{\gamma}^{\prime}}), {{m}^{\prime}} \rangle =
	\sum_{m,s} \langle ( j(\gamma), m )( 1,s) | j({{\gamma}^{\prime}} ),
	{{m}^{\prime}} \rangle \; \left( | s \rangle \otimes | \gamma
	\rangle \right) . 
\]
Inserting this into Eq.~(\ref{eq:proj}) we find that
\[
| {{\gamma}^{\prime}}, {{m}^{\prime}} \rangle = \sum_{s,(\gamma,m) } A^{
({{\gamma}^{\prime}}, {{m}^{\prime}}),(\gamma,m)}[s] ( | s \rangle \otimes |
\gamma, m \rangle ),  
\] 
where
\[
A^{ ({{\gamma}^{\prime}}, {{m}^{\prime}} ),(\gamma,m)}[s] =
	 P^{{{\gamma}^{\prime}}, \gamma } \langle (j(\gamma),m) \; ( 1,s ) |
	 j({{\gamma}^{\prime}} ),{{m}^{\prime}} \rangle . 
\]
Thus, although the projection matrices $ A $ contain a total of $ 3 \times
12 \times 12 $ numbers, they are in fact generated by the relatively few
degrees of freedom available in $ P^{{{\gamma}^{\prime}}, \gamma } $.

For this case with twelve basis states there are naively 16 parameters in
$P^{{{\gamma}^{\prime}},\gamma}$.  Demanding normalization of all basis
states, $\langle \gamma_1, m_1 | \gamma_2, m_2 \rangle = \delta_{\gamma_1,
\gamma_2} \delta_{m_1, m_2} $, yields the condition that the diagonal
elements of $P^T P$ are all $1$, where the superscript $T$ denotes
transpose. This gives four constraints. (Cf.\ Eq.~(\ref{eq:ortho}).) Then
the basis states of the two spin-1/2 and the two spin-3/2 must be
orthogonal, yielding the condition $( P^T P )^{\gamma_1, \gamma_2} = 0$,
whenever $j(\gamma_1) = j(\gamma_2)$ with $\gamma_1 \neq \gamma_2$. This
gives two more constraints. The spin-1/2 basis states are automatically
orthogonal to the spin-3/2 states. Finally, a unitary transformation can mix
the two spin-1/2 and likewise the two spin-3/2. Without loss of generality
we can fix this freedom, yielding two more constraints. We thus end up with
only eight free parameters \cite{Kennedy92hids} in
$P^{{{\gamma}^{\prime}},\gamma}$. In the simpler case of saving only six
basis states, only two free parameters are available by similar arguments.

With only a few free parameters we can use a variational principle for the
energy to determine these.  At this point it is clear that the DMRG plays no
essential role in the construction aside from providing a guide to which
representations to keep. Even this choice could be done variationally.


\section{The set of states $|Q)$}
\label{sec:q}

\subsection{The ground state ansatz}
To do the variational calculation we need an expression for the energy. As
an ansatz for the ground state wave function we take the translationally
invariant state $Q=\openone$ from Eq.~(\ref{eq:qstate}) which we denote by
$| 1 \rangle $. Thus
\begin{equation}
| 1 \rangle \equiv \sum_{\{ s_j \} } { \, \mbox{tr} \left( \;   A[s_n]
\ldots A[s_1]  \; \right) } | s_n \ldots s_1 \rangle . \label{eq:gnddef} 
\end{equation}
Note that although it is not explicitly written out, $|1\rangle$ has a
definite number of lattice sites $n$. We note that $ \langle 1 | 1 \rangle =
1 $ due to Eq.~(\ref{eq:ortho}).  For the AKLT model \cite{AKLT88val}
($\beta = -1/3$) our ground state ansatz is exact as are the ``matrix
product'' states of Ref.~\onlinecite{Accardi81top,Fannes89exa,Klumper93mat}.

The expectation value of an operator $h$, e.g.\ energy or correlation
function, in this state is given by
\begin{equation}
\langle 1 | h | 1 \rangle = \sum_{\{ s_j \}, \{ {{s}^{\prime}}_j \} } {
\, \mbox{tr} \left( \; A^*[{{s_n}^{\prime}}] \ldots A^*[{{s_1}^{\prime}}]
\; \right) } { \, \mbox{tr} \left( \;   A[s_n] \ldots A[s_1]  \; \right) }
\langle {{s_n}^{\prime}} \ldots {{s_1}^{\prime}} | h | s_n \ldots s_1
\rangle . \label{eq:qopq}  
\end{equation}

To write this expression in a simpler form we define the tensor product
matrix $(B \otimes C)$ by $(B \otimes C)^{(\alpha, \beta), (\tau, \nu)} =
B^{\alpha, \tau} C^{\beta, \nu}$. We will in the rest of the article
interchangeably use ordinary matrix indices $\alpha, \beta$ and composite
indices $(\alpha, \beta)$, where composite indices are written with a
parenthesis around them. This means that we can write a $m \times m$
``matrix'' $A$ as either a matrix $A^{\alpha, \beta}$ or as a $m^2$ vector
$A^{(\alpha, \beta)}$. When the indices are not explicitly written out, the
matrix or vector character of the symbol is assumed to be clear from the
context.  We now use the trace and matrix product identities ${ \, \mbox{tr}
\left( \; B \; \right) } { \, \mbox{tr} \left( \; C \; \right) } = { \,
\mbox{tr} \left( \; B \otimes C \; \right) }$ and $(B C D) \otimes (E F G) =
(B \otimes E) (C \otimes F) (D \otimes G)$ to find
\begin{equation}
\langle 1|h|1 \rangle = \sum_{\{ s_j \}, \{ {{s_j}^{\prime}} \} } { \,
\mbox{tr} \left( \; (A^*[{{s_n}^{\prime}}] \otimes A[s_n]) \ldots
(A^*[{{s_1}^{\prime}}] \otimes A[s_1])  \; \right) } \langle
{{s_n}^{\prime}} \ldots {{s_1}^{\prime}} | h | s_n \ldots s_1 \rangle
. \label{eq:1h1}   
\end{equation}

To write this in a more compact form we define a mapping $\widehat{M}$ from
$3 \times 3$ spin matrices $M$ to $m^2 \times m^2$ matrices $\widehat{M}$ by
\begin{equation}
\widehat{M} \equiv \sum_{{{s}^{\prime}},s} M_{{{s}^{\prime}},s} \; (
A^*[{{s}^{\prime}}] \otimes A[s] ) . \label{eq:hatdef} 
\end{equation}
We denote by $S \equiv (S^x, S^y, S^z)$ the spin-1 representation of total
spin and thus by $\hat{S} \equiv (\hat{S}^x, \hat{S}^y, \hat{S}^z )$ the
``hat'' mapping of the $3 \times 3$ spin matrices $S$. By $\widehat{1}$ we
denote the ``hat'' mapping of the $3 \times 3$ identity matrix. We now see
from Eq.~(\ref{eq:1h1}) that the norm and the expectation value of the spin
at the site $j$ is given by
\begin{eqnarray*}
\langle 1 | 1 \rangle & = & { \, \mbox{tr} \left( \;   \hat{{\sl 1 }}^n  \;
\right) } \\ 
\langle 1|S_j|1 \rangle & = & { \, \mbox{tr} \left( \;   \hat{{\sl 1
}}^{n-1} \hat{S}  \; \right) }, 
\end{eqnarray*}
where we in the last equation have used the cyclicity of the trace. Other
expectation values are also easily obtained. Since we can factorize matrix
elements like
\begin{eqnarray*}
\langle {{s}^{\prime}}_j, {{s}^{\prime}}_i | {\bf S}_i \cdot {\bf S}_j |
s_j, s_i \rangle & \equiv & ( {\bf S}_i \cdot {\bf S}_j )_{{{s}^{\prime}}_j,
{{s}^{\prime}}_i, s_j, s_i } \\ 
& = & ({\bf S})_{{{s}^{\prime}}_i, s_i} \cdot ({\bf S})_{{{s}^{\prime}}_j,
s_j}, 
\end{eqnarray*}
we find that expectation values of energy and spin-spin correlation function
are given by
\begin{eqnarray}
\langle 1|{\bf S}_j \cdot {\bf S}_{j+1}|1 \rangle & = & { \, \mbox{tr}
\left( \;   \hat{{\sl 1 }}^{n-2} \hat{S} \hat{S}  \; \right) } \nonumber \\
\langle 1| {\bf S}_j \cdot {\bf S}_{j+l} |1 \rangle & = & { \, \mbox{tr}
\left( \;   \hat{{\sl 1 }}^{n-l-1} \hat{S} \hat{{\sl 1 }}^{l-1} \hat{S}  \;
\right) } . \label{eq:sscorr}  
\end{eqnarray}
Similar formulas have also been derived by Fannes {\it et.al}
\cite{Fannes89exa}. 

A more complicated operator, like the biquadratic term $({\bf S}_i \cdot
{\bf S}_j)^2$, does not factorize as neatly and we cannot write the
expectation value in such a nice form as above.  For these cases we have to
replace the term $\hat{S} \hat{{\sl 1 }}^{l-1} \hat{S}$ inside the trace in
Eq.~(\ref{eq:sscorr}) by the more complicated expression
\begin{equation}
\sum_{{{s}^{\prime}}_j, {{s}^{\prime}}_i, s_j, s_i} \langle
{{s}^{\prime}}_j, {{s}^{\prime}}_i | ( {\bf S}_i \cdot {\bf S}_j )^2 | s_j,
s_i \rangle \; (A[{{s}^{\prime}}_i] \otimes A[s_i]) \hat{{\sl 1 }}^{l-1}
(A[{{s}^{\prime}}_j] \otimes A[s_j]) . \label{eq:complhat} 
\end{equation}
In order not to make the equations unreadable by crowding them with indices
we will in the rest of this chapter only present formulas for the ordinary
bilinear Heisenberg Hamiltonian ($\beta = 0$ in Eq.~(\ref{eq:heis})). An
interested reader can then generalize the formulas to include the
biquadratic term, without any fundamental difficulties.

An important quantity is the string correlation function
\cite{Nijs89pre} defined by
\begin{equation}
g(l) = \langle S_0^z \left( \prod_{j=1}^{l-1} e^{ i \pi S_j^z } \right)
 S_l^z \rangle . \label{eq:string}
\end{equation}
Although the spin-1 chain does not have long range N\'eel order, it is
believed to have a hidden long range order that is characterized by the
string correlation function. In our ground state $|1\rangle$ it is easy to
show that it is given by
\begin{displaymath}
g(l) = { \, \mbox{tr} \left( \;   \hat{{\sl 1 }}^{n-l-1} \hat{S^z} ({
\widehat{ e^{ i \pi S_z } } })^{l-1} \hat{S^z}  \; \right) }. 
\end{displaymath}

We note that the spectrum of correlation lengths, i.e.\ the collection of
all possible exponential decay lengths $\xi$ of correlation functions of the
form $\langle {\cal O}_1(x) {\cal O}_2(y) \rangle \propto e^{ -|x-y|/ \xi
}$, is determined by the eigenvalues of $\hat{1}$.  One can show that
$\hat{1}$ is guaranteed to have an eigenvalue of $1$ due to
Eq.~(\ref{eq:ortho}), and numerically we find that all other eigenvalues
have absolute value less than $1$. It is however not true that the
eigenvalue $1$ will always dominate.  If each of the rows of $\hat{\cal
O}_1$ or each of the columns of $\hat{\cal O}_2$ is orthogonal to this
particular eigenvector, another eigenvalue will determine the correlation
length. Thus, the correlation length $\xi$ is given by
\begin{equation}
\xi = -\frac{1}{\log{x}}, \label{eq:corrlog}
\end{equation}
where $x$ is the largest eigenvalue of $\hat{1}$ not orthogonal to the
operator. Since the rows and columns of the spin operator $\hat{S}$ turn out
to be orthogonal to the eigenvalue $1$ while the next leading eigenvalue is
not, the next largest eigenvalue will determine the decay of spin
correlations. The string operator $e^{ i \pi S_j^z }$ of
Eq.~(\ref{eq:string}) turns out to have the same eigenvalue spectrum as
$\hat{1}$. This time however, the eigenvalue $1$ of $\hat{{\sl 1 }}$ is not
orthogonal to $ { \widehat{ e^{ i \pi S_z } } } $, giving the long range
string correlations.

A possible problem with the construction of the projection operator is that
parity is not built into the construction of the ground state since the
projectors operate from the left to the right. There is therefore the
possibility that parity is violated in the ground state $|1 \rangle $.  We
now investigate this possibility and show how parity is maintained.

Let $ {\cal P } $ be the parity operator. We thus have
\begin{eqnarray*}
{\cal P } | 1 \rangle & = & \sum_{\{s_j\}} { \, \mbox{tr} \left( \;   A[s_n]
\ldots  A[s_1 ]  \; \right) } {\cal P } | s_n \ldots s_1 \rangle \\  
& = & \sum_{\{s_j\}} { \, \mbox{tr} \left( \;   A[s_n] \ldots A[s_1]  \;
\right) } | s_1 \ldots s_n \rangle . 
\end{eqnarray*}
Suppose now that there exists an invertible $m \times m$ matrix $
Q_{{\cal P }} $ such that 
\begin{equation}
Q_{{\cal P }} A[s] = sign[ {\cal P }]  ( A[s])^T Q_{{\cal P }},
\label{eq:pardef} 
\end{equation}
where $ A^T $ denotes transpose and $sign[ {\cal P } ]$ is a proportionality
constant that will be seen to be the eigenvalue of the parity operator. Then
it follows that
\begin{eqnarray*}
{\cal P } | 1 \rangle & = & \sum_{\{s_j\}} { \, \mbox{tr} \left( \;
Q_{{\cal P }}^{-1} Q_{{\cal P }} A[s_n] \ldots A[s_1]  \; \right) } | s_1
\ldots s_n \rangle \\  
& = & sign[{\cal P }]^n \sum_{\{s_j\}} { \, \mbox{tr} \left( \;   A^T [s_n]
\ldots A^T [s_1 ]  \; \right) } | s_1 \ldots s_n \rangle \\ 
& = & sign[{\cal P }]^n \sum_{\{s_j\}} { \, \mbox{tr} \left( \;   A[s_1]
\ldots A[s_n ]  \; \right) } | s_1 \ldots s_n \rangle \\ 
& = & sign[{\cal P }]^n | 1 \rangle .
\end{eqnarray*}
Thus, for the ground state to have definite parity, it is sufficient that
such a $ Q_{{\cal P }} $ exists.  How do we find this matrix, if it exists?
We multiply both sides of the defining relation Eq.~(\ref{eq:pardef}) by $
A^{\dagger}[s] $ and sum over $ s $. Using Eq.~(\ref{eq:ortho}) we find that
\begin{eqnarray*}
Q_{{\cal P }}^{\alpha,\beta} & = & sign[{\cal P }] \sum_{s}
(A^T[s])^{\alpha,\tau} Q_{{\cal P }}^{\tau,\nu} (A^T[s])^{\nu,\beta} \\    
& = & sign[{\cal P }] \left(  \sum_{s}(A^T[s])^{\alpha,\tau} A^{\beta,\nu}
[s] \right) Q_{{\cal P }}^{\tau,\nu} \\
& = & sign[{\cal P }] \sum_{s} ( A^T[s] \otimes A[s]
)^{(\alpha,\beta),(\tau,\nu)} Q_{{\cal P }}^{\tau,\nu} .  
\end{eqnarray*}
Thus, $ Q_{{\cal P }} $, if it exists, is the eigenvector of the matrix $
\sum_{s} ( A^T[s] \otimes A[s] ) $ with eigenvalue $ \pm 1 $. It is
straightforward to numerically obtain the eigenvalue spectrum of this
operator, and in the cases that we have looked at, this parity operator
exists.

\subsection{The general state $|Q)$ }
\label{sec:edge}
We now analyze the set of states $|Q)_n$, for general $Q$. These states can
be interpreted as states homogeneous in the bulk but with nonuniformity near
the boundary.

To calculate the norm we use the same trace and tensor product identities as
when deriving Eq.~(\ref{eq:1h1}). We find that
\begin{eqnarray}
( {{Q}^{\prime}} | Q )_n & = & \sum_{ \{ s_j \} } { \, \mbox{tr} \left( \;
({{Q}^{\prime}})^* A^*[s_n] \ldots A^*[s_1]  \; \right) } { \, \mbox{tr}
\left( \;   Q A[s_n] \ldots A[s_1]  \; \right) } \nonumber \\ 
& = & { \, \mbox{tr} \left( \;   ( {{Q}^{\prime}}^* \otimes Q ) \; \hat{1}^n
\; \right) } . \label{eq:qtrace} 
\end{eqnarray}
We can rewrite this trace as ordinary matrix products. To do this we first
define the generalized transpose $M^{T_{p_1, p_2, p_3, p_4}}$ of a matrix
$M$ by
\begin{equation}
(M^{T_{p_1, p_2, p_3, p_4}})^{(\alpha_1, \alpha_2), (\alpha_3, \alpha_4)} = 
M^{(\alpha_{p_1}, \alpha_{p_2}), (\alpha_{p_3}, \alpha_{p_4})} ,
\label{eq:gentr} 
\end{equation}
where $\{ p_1, p_2, p_3, p_4 \}$ is a permutation of $\{ 1, 2, 3, 4 \}$. We 
also define 
a tilde operator $\widetilde{M}$ by the formula
\begin{equation}
{ \, \mbox{tr} \left( \;   ( {{Q}^{\prime}} \otimes Q ) M  \; \right) } =
\sum_{ {{\alpha}^{\prime}}, {{\beta}^{\prime}}, \alpha, \beta } (
{{Q}^{\prime}} )^{{{\alpha}^{\prime}}, {{\beta}^{\prime}}} ( \widetilde{M}
)^{({{\alpha}^{\prime}}, {{\beta}^{\prime}}), (\alpha, \beta)} Q^{\alpha,
\beta} , \label{eq:tildedef} 
\end{equation}
so that the tilde operator effectively generates the matrix corresponding to
the inner product of ${{Q}^{\prime}}$ and $Q$ with $M$. One finds by writing
out Eq.~(\ref{eq:tildedef}) in components that
\begin{displaymath}
\widetilde{M} = M^{T_{3142}}.
\end{displaymath}
Hence
\begin{equation}
( ({{Q}^{\prime}})^{{{\alpha}^{\prime}}, {{\beta}^{\prime}}} | Q^{\alpha,
\beta} )_n = {{Q}^{\prime}}^{({{\alpha}^{\prime}}, {{\beta}^{\prime}})}
G(n)^{({{\alpha}^{\prime}}, {{\beta}^{\prime}}), (\alpha, \beta)}
Q^{(\alpha, \beta)} , \label{eq:qnorm} 
\end{equation}
with 
\begin{equation}
G(n) = \widetilde{ ( \hat{1}^n ) } . \label{eq:gdef}
\end{equation}
The nice thing about Eq.~(\ref{eq:qnorm}) and (\ref{eq:gdef}) is that we
have effectively turned the computation of the trace in
Eq.~(\ref{eq:qtrace}) for all $Q$ and ${{Q}^{\prime}}$ into a matrix inner
product between $Q$, ${{Q}^{\prime}}$ and a single $m^2 \times m^2$ matrix
$G$, independent of $Q$ and ${{Q}^{\prime}}$. Note that on the right side in
Eq.~(\ref{eq:qnorm}) we write $Q$ and ${{Q}^{\prime}}$ as vectors of length
$m^2$.

Similarly we can compute the expectation value of the Heisenberg Hamiltonian
defined in Eq.~(\ref{eq:heis}) with $\beta = 0$ as
\begin{equation}
( {{Q}^{\prime}} | H_{op} | Q )_n = \sum_{i=0}^{n-2} { \, \mbox{tr} \left(
\;   ({{Q}^{\prime}} \otimes Q) \hat{1}^i \hat{S} \hat{S} \hat{1}^{n-2-i}
\; \right) } , \label{eq:qham}  
\end{equation}
where $\hat{S}$ denotes the hat mapping in Eq.~(\ref{eq:hatdef}) of the
spin-1 matrices. The $z$-component of total spin, $(S^z_T)_{op} = \sum_{i}
S_i^z $ is given by
\begin{equation}
( {{Q}^{\prime}} | (S^z_T)_{op} | Q )_n = \sum_{i=0}^{n-1} { \, \mbox{tr}
\left( \;   ({{Q}^{\prime}} \otimes Q) \hat{1}^i \hat{S^z} \hat{1}^{n-1-i}
\; \right) } . \label{eq:qsz} 
\end{equation}
If we have a more complicated Hamiltonian, like Eq.~(\ref{eq:heis}) with
$\beta \neq 0$, the Hamiltonian matrices $\hat{S} \hat{S}$ in
Eq.~(\ref{eq:qham}) must be replaced by an expression similar to
Eq.~(\ref{eq:complhat}).  As we did with the norm in Eq.~(\ref{eq:qnorm}) we
can rewrite Eq.~(\ref{eq:qham}) and (\ref{eq:qsz}) as matrix products by
putting the summations inside the traces and by using the tilde
transformation of Eq.~(\ref{eq:tildedef}), yielding
\begin{eqnarray}
( {{Q}^{\prime}} | H_{op} | Q )_n & = & {{Q}^{\prime}} H(n) Q \label{eq:qhq}
\\ 
( {{Q}^{\prime}} | (S^z_T)_{op} | Q )_n & = & {{Q}^{\prime}} S_T^z(n) Q , 
\end{eqnarray}
where 
\begin{eqnarray}
H(n) & = & \widetilde{\sum_{i=0}^{n-2}} \left( \hat{1}^i \hat{S}
\hat{S} \hat{1}^{n-2-i} \right) \label{eq:qhop} \\
S^z_T(n) & = & \widetilde{\sum_{i=0}^{n-1}} \left( \hat{1}^i \hat{S^z}  
\hat{1}^{n-2-i} \right) \label{eq:qsop} ,
\end{eqnarray}
where the tilde symbols indicate that the transformation in
Eq.~(\ref{eq:tildedef}) should be performed on the whole sum.

In Eq.~(\ref{eq:qnorm}) we determined the expression $( {{Q}^{\prime}} | Q
)_n = {{Q}^{\prime}} G(n) Q $ for the inner product of the states $|Q)_n$
and in Eq.~(\ref{eq:qhq}) we found $({{Q}^{\prime}}|H_{op}|Q) =
{{Q}^{\prime}} H(n) Q$. Since $G$ turns out not to be proportional to the
identity matrix, we see that the naive basis states, i.e.\ the states
$(Q_{i,j})^{\alpha, \beta} \equiv \delta_{i,\alpha} \delta_{j,\beta},$ with
$i = 1 \ldots m$ and $j = 1 \ldots m $, are not orthonormal. It is not only
convenient to have an orthonormal set of states, we also want them to be
eigenstates of the Hamiltonian.  The energy of the state $|Q)_n$ defined by
the $m \times m$ matrix $Q$ is given by
\begin{equation}
E_Q(n) = \frac{Q H(n) Q}{Q G(n) Q}. \label{eq:hdivg}
\end{equation}
The eigenvalue equation we have to solve is thus
\begin{equation}
H(n) Q = E_Q G(n) Q . \label{eq:eigeq}
\end{equation}

We will now construct a set of states that are orthonormal and satisfies
Eq.~(\ref{eq:eigeq}).  Since $G$ is Hermitian we can define a unitary matrix
$V$ by the transformation that diagonalizes $G$:
\begin{equation}
V^\dagger G V = D_G , \label{eq:vgv}
\end{equation}
where $D_G$ is a diagonal matrix. We now define 
\begin{displaymath}
u = V (D_G)^{-1/2}
\end{displaymath}
so that $ u^\dagger G u = \openone $, the identity matrix. We also define
\begin{eqnarray}
h & = & u^\dagger H u \label{eq:uhu} \\
s_z & = & u^\dagger S_T^z u , \label{eq:usu}
\end{eqnarray}
with $H$ and $S_T^z$ from Eq.~(\ref{eq:qhop}) and Eq.~(\ref{eq:qsop}). It
can be verified that $\left[ h, s_z \right] = 0$ so that both total spin and
the energy can be diagonalized simultaneously. Numerically we diagonalize $h
+ \epsilon s_z$ where $\epsilon$ is a small number, so that $w^\dagger ( h +
\epsilon s_z ) w = E + \epsilon s_z$ is diagonal and we find that both $h$
and $s_z$ are thereby diagonalized by
\begin{eqnarray}
w^\dagger h w & = & E \label{eq:whw} \\
w^\dagger s_z w & = & m_z , \label{eq:wsw}
\end{eqnarray}
with $w^\dagger w = \openone$ and where $E$ and $m_z$ are diagonal matrices
containing the energy eigenvalues and the eigenvalues of total spin
respectively. Putting Eq.~(\ref{eq:uhu}) and (\ref{eq:usu}) into
Eq.~(\ref{eq:whw}) and (\ref{eq:wsw}) we see that
\begin{eqnarray}
\sum_{\alpha, \beta} \left( ( u w )^\dagger \right)_{{{\gamma}^{\prime}},
\alpha} H_{\alpha, \beta} ( u w )_{\beta, \gamma} & = & E_\gamma
\delta_{{{\gamma}^{\prime}}, \gamma}  \label{eq:uwhwu} \\ 
\sum_{\alpha, \beta} \left( ( u w )^\dagger \right)_{{{\gamma}^{\prime}},
\alpha} G_{\alpha, \beta} ( u w )_{\beta, \gamma} & = &
\delta_{{{\gamma}^{\prime}}, \gamma} , \label{eq:uwgwu} 
\end{eqnarray}
where $\alpha, \beta, \gamma$ and ${{\gamma}^{\prime}}$ are matrix indices.
Thus, the columns of $(uw)$ contain the orthonormal eigenvectors of $H(n)$
and $S_T^z(n)$. Combining Eq.~(\ref{eq:uwhwu}) and (\ref{eq:uwgwu}) we find
\begin{displaymath}
\sum_{\beta} H_{\alpha, \beta} (uw)_{\beta, \gamma} = E_\gamma
G_{\alpha, \beta} (uw)_{\beta, \gamma} .
\end{displaymath}
Hence the matrices
\begin{equation}
( Q_\gamma )_{\alpha, \beta} = ( u w )_{(\alpha, \beta), \gamma} , 
\label{eq:quw} 
\end{equation}
where $ Q_\gamma $ are $m^2 $ $ m \times m$ matrices, are orthogonal with
respect to $G$ and are simultaneous eigenstates of $H$ and $S_T^z$. We
therefore define the orthonormal set of states $| \gamma \rangle $ we were
looking for by
\begin{displaymath}
| \gamma \rangle \equiv | Q_\gamma ) .
\end{displaymath}
To summarize, we finally have
\begin{eqnarray*}
\langle {{\gamma}^{\prime}} | H_{op} | \gamma \rangle  & = & E_\gamma
\delta_{{{\gamma}^{\prime}}, \gamma} \\
\langle {{\gamma}^{\prime}} | (S^z_T)_{op} | \gamma \rangle & = &
(m_z)_\gamma \delta_{{{\gamma}^{\prime}}, \gamma} \\
\langle {{\gamma}^{\prime}} | \gamma \rangle & = &
\delta_{{{\gamma}^{\prime}}, \gamma} .  
\end{eqnarray*}
The states $| \gamma \rangle $ form a natural basis for describing edge
states in finite size calculations, a feature which is not further explored
in this manuscript.  Nevertheless, we will benefit from this derivation in
the next section were a set of Bloch states are defined in a similar manner.

\section{Bloch states}
\label{sec:bloch}
We now leave the orthonormal boundary states $|Q_\gamma)$ and impose
periodic boundary conditions on the Hamiltonian in Eq.~(\ref{eq:heis}). We
return to the states $|Q)_n$ as defined in Eq.~(\ref{eq:qstate}), where $Q$
is a general $m \times m$ matrix, to make an ansatz for the low lying
excited states. For a translationally invariant system we can define our
states to be Bloch states. A reasonable ansatz for a Bloch state $|Q,k)_n$
defined by a matrix $Q$ and a momentum $k$ is given by
\begin{equation}
|Q,k)_n \equiv \sum_{ \{ s_j \} } \sum_j e^{ijk} { \, \mbox{tr} \left( \;
A[s_n] \ldots A[s_{j+1}] Q A[s_j] \ldots A[s_1]  \; \right) } | s_n \ldots
s_1 \rangle . \label{eq:bloch} 
\end{equation}
This wavefunction can be viewed as the ground state $| 1 \rangle$ with a
disturbance $Q$ introduced at some site, and then letting the disturbance
run over all sites to form a state with a definite momentum. In this way we
get a single ``particle'' excitation.

As was done for the boundary states in Sec.~\ref{sec:q} we now derive
expressions for expectation values of operators in the states $|Q,k)_n$. The
calculations are more tedious and we have therefore put the details in
App.~\ref{app:expval} - \ref{app:hampole}. The resulting expressions are in
principle similar to the ones we obtained for the boundary states, e.g.\
Eq.~(\ref{eq:qham}). For the norm we find
\begin{equation}
( {{Q}^{\prime}}, k | Q, k ) = n \; { \, \mbox{tr} \left( \;
({{Q}^{\prime}} \otimes \openone ) \sum_{j=0}^{n-1} e^{ijk} \hat{{\sl 1
}}^{n-j} (\openone \otimes Q) \hat{{\sl 1 }}^{j}  \; \right) } ,
\label{eq:qkqk}  
\end{equation}
with a similar, but more complicated, expressions for the Hamiltonian and
for the $z$-component of total spin. The results can be found in
Eq.~(\ref{eq:qkstot}) and (\ref{eq:qkham}).  We see that the general
structure of all these matrix elements are that they consist of traces with
a convolution sum over matrix products inside each trace.
For finite length chains, the sums in these expressions as well as those in
Eq.~(\ref{eq:qham}) and (\ref{eq:qsz}) can be expediently calculated by a
recursive scheme for the case when $n$ is a power of two. These recursive
formulas are derived in App.~\ref{app:recsum}.

One can also calculate the norm and Hamiltonian matrices, $G(k,n)$ and
$H(k,n)$, defined through the formula
\begin{eqnarray}
({{Q}^{\prime}},k | H_{op} | Q,k)_n & = & n \; {{Q}^{\prime}} H(k,n) Q
\label{eq:qhkq} \\ 
({{Q}^{\prime}},k | Q,k)_n & = & n \; {{Q}^{\prime}} G(k,n) Q ,
\label{eq:qgkq}  
\end{eqnarray}
similar to $H(n)$ and $G(n)$ in Eq.~(\ref{eq:qnorm}) and (\ref{eq:qhq}) for
the boundary states. This time they will however depend on $k$ as well as on
$n$. A matrix $S_T^z(k,n)$ representing the $z$-component of total spin can
be defined analogously. The principles for calculating these matrices are
the same as for the boundary states, i.e.\ one uses the tilde transformation
of Eq.~(\ref{eq:tildedef}). Due to the number of terms in the expression for
the expectation values it is numerically cumbersome for finite length
chains.

There is however an elegant way to extract the leading behavior of $H(k,n)$
and $G(k,n)$ as $n \rightarrow \infty$. The details of these calculations
can be found in App.~\ref{app:pole} and \ref{app:hampole}. In this section
we will only give a brief summary of the method and the results.  Let us
first define the $z$-transform (sometimes called a discrete Laplace
transform) of a series $\{ a_n \}_{n=0}^\infty$ by $F(\lambda) =
\sum_{n=0}^\infty a_n e^{-n \lambda}$. Let us now denote the sum inside the
trace in Eq.~(\ref{eq:qkqk}) by $S_n$, so that
\[
S_n = \sum_{j=0}^{n-1} e^{ijk} \hat{{\sl 1 }}^{n-j} (\openone \otimes Q)
\hat{{\sl 1 }}^j .  
\]
We now define a series $\{ S_n \}_{n=0}^\infty$, and take the $z$-transform
of this sequence. By examining the analytical structure of the transformed
series we are able to extract the leading behavior of the sum $S_n$, as $n
\rightarrow \infty$. In this way we get the asymptotic form of the norm in
the limit of large $n$. This procedure is then applied to all sums in the
expressions for the matrix elements. In App.~\ref{app:pole}, the
$z$-transform of a general sum is taken and its large $n$ behavior is
extracted. In App.~\ref{app:hampole} we apply the results of
App.~\ref{app:pole} to the expressions for the expectation values derived in
App.~\ref{app:expval}. This whole procedure finally results in the
asymptotic forms
\begin{eqnarray}
H(k,n) & = & n^2 H_2(k) + n H_1(k) + H_0(k) + {\cal O}(z)^n
\label{eq:hasy}\\ 
G(k,n) & = & n G_1(k) + G_0(k) + {\cal O}(z)^n \label{eq:gasy} ,
\end{eqnarray}
with $H(k,n)$ and $G(k,n)$ as defined in Eq.~(\ref{eq:qhkq}) and
(\ref{eq:qgkq}). Here $z$ represents the next leading eigenvalue of
$\hat{{\sl 1 }}$ and we find numerically that $|z| \approx 0.8$. There are
thus very small corrections to the asymptotic form. We also find that $H_2$
and $G_1$ are nonvanishing only when the momentum $k$ is zero.  The
eigenvalue equation which must be solved is
\[
H(k,n) Q(k,n) = E(k,n) G(k,n) Q(k,n) ,
\]
where $Q(k,n)$ is an $m^2$-dimensional vector.
For $k \neq 0$ we thus have
\begin{equation}
\left( n H_1(k) + H_0(k) \right) Q(k,n) = ( n E_0 + \Delta_k(n) ) G_0(k) 
Q(k,n) , \label{eq:hqkn}
\end{equation}
where $E_0$ is the ground state energy per site and $\Delta_k(n)$ is the
excitation energy.  $E_0$ denotes the ground state energy per site in the
limit $n \rightarrow \infty$, and is therefore independent of $n$. Since we
are interested in the solutions to Eq.~(\ref{eq:hqkn}) when $n \rightarrow
\infty$ we assume $Q$ and $\Delta_k$ to be independent of $n$ and we thus
need to solve the simultaneous equations
\begin{eqnarray}
H_1(k) Q(k) & = & E_0 G_0(k) Q(k) \label{eq:ev1} \\
H_0(k) Q(k) & = & \Delta_k G_0(k) Q(k) \label{eq:ev2} .
\end{eqnarray}
Solving Eq.~(\ref{eq:ev2}) yields a set of eigenstates $Q(k)$ and
eigenvalues $\Delta_k$ for each $k$. These eigenstates have to be
simultaneous eigenstates to Eq.~(\ref{eq:ev1}) with the $k$-independent
eigenvalue $E_0$. This is in general impossible, unless $H_1 \propto G_0$,
as indeed happens.  We thus recover $E_0$ by the proportionality constant 
\[
H_1(k) = E_0 G_0(k) .
\]
The excitation spectrum is then given by the single eigenvalue equation  
\begin{equation}
H_0(k) Q(k) = \Delta_k G_0(k) Q(k) . \label{eq:exceig}
\end{equation}
Similar formulas can be obtained for $k=0$. Note that Eq.~(\ref{eq:exceig})
is an eigenvalue equation for the excitation spectrum which makes no
explicit reference to a ground state. The ground state enters however
implicitly in the parameters in $A[s]$ on which $H(k,n)$ and $G(k,n)$
depends.

An asymptotic form for the $z$-component of total spin, $S_T^z(k,n)$,
similar to the form for $H(k,n)$, containing terms up to order $n^2$ is also
derived in App.~\ref{app:hampole}. Numerically we find however, that the
only nonvanishing term in $S_T^z(k,n)$, for any momentum $k$, is the
constant term $S_0^z(k)$.

How do we find the orthonormal set of states $Q(k)$ for a particular $k$
from the eigenstate equation in Eq.~(\ref{eq:exceig})?  We can in principle
take over the discussion of the boundary states $|Q)$ from
Sec.~\ref{sec:q}. The only slight problem that enters here is that $G_0(k)$
is singular for $k \neq 0$, that is, the nullspace of $G_0(k)$ is
nonvanishing. In order to find the inverse of $G_0(k)$ the nullspace must be
excluded from the Hilbert space. We do this numerically using singular value
decomposition. Once this has been done we can simply take over
equations~(\ref{eq:vgv})-(\ref{eq:quw}). In this case we identify $H_0(k)$
with $H$, $G_0(k)$ with $G$ and $\Delta_k$ with $E$. We diagonalize $G_0(k)$
with $V_k$ so that $V_k^\dagger G_0(k) V_k = D_G(k)$ is diagonal and define
$u_k = V_k (D_G(k))^{-1/2}$. Then we diagonalize $h_k + \epsilon s_k^z$,
where $h_k = u_k^\dagger H_0(k) u_k$ and $s_k^z = u_k^\dagger S_0^z(k) u_k$
so that $w_k^\dagger (h_k + \epsilon s_k^z ) w_k = \Delta_k + \epsilon
m_k^z$ is diagonal. We then find
\[
H_0(k) Q_\gamma (k) = \Delta_{k,\gamma} G_0(k) Q_\gamma (k) , 
\]
where
\[
(Q_\gamma (k))_{\alpha,\beta} \equiv (u_k w_k)_{(\alpha,\beta),\gamma}
\]
are $m \times m$ matrices labeled by $\gamma$, orthogonal with respect to
$G_0(k)$ and simultaneous eigenstates of $H_0(k)$ and $S_0(k)$. There are
less than $m^2$ eigenvectors $Q_\gamma (k)$ for $k \neq 0$ due to the
nonvanishing nullspace of $G_0(k)$. However, probably only a few of the
lowest lying energy eigenstates $Q_\gamma (k)$ are reasonable estimates of
true excited states. Finally, we can write for the orthonormalized states
$|\gamma, k \rangle$, defined by the matrices $Q_\gamma (k)$
\begin{equation}
|\gamma, k \rangle \equiv | Q_\gamma(k), k ) . \label{eq:gammak}
\end{equation}
Because states with different values of $k$ are guaranteed to be orthogonal,
we find
\[
\langle {{\gamma}^{\prime}}, {{k}^{\prime}} | \gamma , k \rangle =
\delta_{{{\gamma}^{\prime}}, \gamma} \delta_{{{k}^{\prime}}, k} .
\]
These represent our ``single magnon'' states. In the next section we
numerically determine these states along with their energy and spin
expectation values.
\section{Results}
\label{sec:results}



We have tested the calculations on the spin-1 Heisenberg chain defined in
Eq.~(\ref{eq:heis}). All computations are done with $m=12$, i.e.\ keeping
the twelve states as discussed in Sec.~\ref{sec:constrofa}. The resultant
eight parameter family of trial ground states (Eq.~(\ref{eq:gnddef})) was
explored. The projection matrices $A[s]$ defining the ground state were
computed by minimizing the energy of the trial ground state. The projection
matrices obtained by this variational technique was found to agree up to
numerical accuracy with the projection operator obtained from similar DMRG
calculations. The result for the lowest energy state for some $\beta$ is
found in Table~\ref{tab:gnd}. The best result known to us for $\beta=0$
comes from DMRG calculations in Ref.~\onlinecite{White93num}. The exact
result at the AKLT point $\beta = -\arctan(1/3)$ can be found in
Ref.~\onlinecite{AKLT88val}. The $\beta = 1$ system was exactly solved using
Bethe ansatz in Ref.~\onlinecite{Takhtajan82pic}. The parity operator of
Eq.~(\ref{eq:pardef}) has been computed in all cases and it is found that
the ground state has parity $(-1)^n$, where $n$ is the number of sites. For
the string order parameter of Eq.~(\ref{eq:string}) we find $g(\infty) =
-0.3759$, whereas best estimates are \cite{White93num} $g(\infty) =
-0.374325096(2)$. We find the next leading eigenvalue of $\hat{{\sl 1 }}$ to
be $-0.777$, giving an asymptotic spin-spin correlation length from
Eq.~(\ref{eq:corrlog}) of $l=3.963$, compared to best estimates
\cite{White93num} of $l=6.03(1)$. We believe that the severe truncation of
our basis to only twelve states has resulted in the asymptotic correlations
being quite poor, although we have verified that intermediate length
spin-spin correlations are consistent with more precise calculations
\cite{Sorensen94hal}.

\begin{figure}
\epsfxsize=\textwidth
\epsfbox{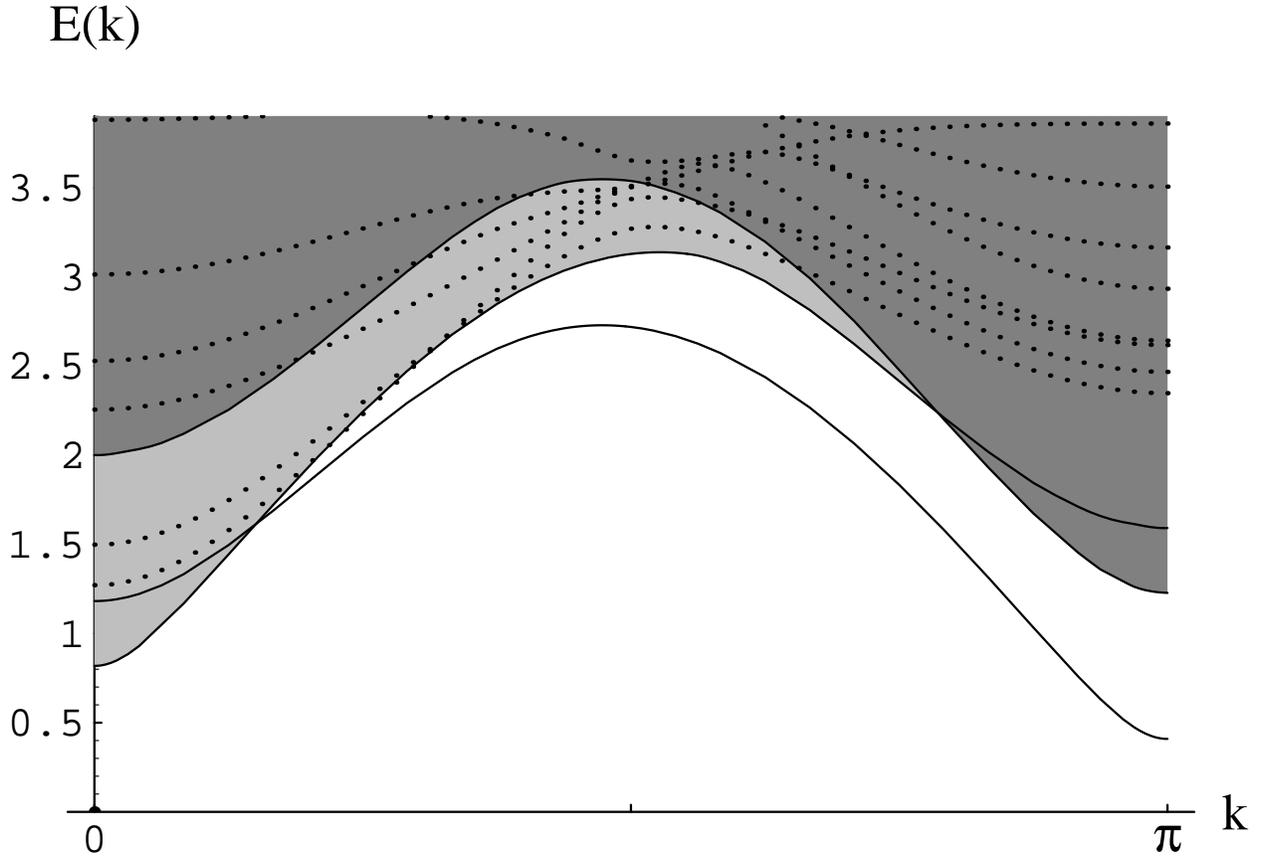}
\caption{The spectrum for $ \beta = 0 $ is shown. The lowest single
	particle triplet is shown as a solid line, with the lightly shaded
	region representing two-particle excitations and the dark region
	three particle excitations. Solid lines define the boundaries to the
	two and three particle continuum.  Dotted lines indicate the
	spectrum of higher energy single-magnon excitations. The spin of
	these dotted excitations are, in order of increasing energy at $ k =
	\pi $: 0,1,2,2,3,1,1,0.}
\label{fig:J2_0}
\end{figure}

An important issue is whether or not Eq.~(\ref{eq:bloch}) is a good ansatz
form for the excitations. We have computed the asymptotic forms when $n
\rightarrow \infty$ for the Hamiltonian and norm matrices defined in
Eq.~(\ref{eq:qhkq}) and (\ref{eq:qgkq}) as well as for the total spin matrix
for different $\beta$ and momenta $k$. The orthonormal eigenstates of
Eq.~(\ref{eq:gammak}) are also determined, giving the single magnon
excitations of our model. The energy and $z$-component of total spin for
each eigenstate are also determined.  A particularly interesting point is
$\beta=0$, the pure Heisenberg model, which has been subject to much
numerical effort. We find the single particle spectrum shown in
Fig.~\ref{fig:J2_0}. The low-lying triplet branch defines the gap
$\Delta_\pi = 0.4094$, which is very good compared to the most accurately
known result \cite{White93num,Sorensen93lar,Sakai90ene} of
$0.410502(1)$. Furthermore, we compute the spin wave velocity $v=2.452$ to
be compared to the calculations in Ref.~\onlinecite{Sorensen94equ}, where
$v=2.49(1)$ was obtained. Clearly we reproduce the single-particle triplet
excitations with high accuracy considering the few number of states in our
basis. Our calculation also yields a detailed spectrum of lowest lying
``single magnon'' excitations shown by dotted lines in
Fig.~\ref{fig:J2_0}. Our second lowest energy excitation at $k=\pi$ is a
singlet shown by a dotted line in Fig.~\ref{fig:J2_0} with $\Delta_\pi
(singlet) = 2.348$. As a function of $k$, the second lowest single-particle
excitation is either a singlet or a \mbox{spin-2} object, as has also been
observed in exact finite size calculations \cite{Oitmaa86cro}. Parity of
each of the elementary excitations is verified by checking the relation
Eq.~(\ref{eq:pardef}) with $Q$ as well as with the matrices $A$. The
boundary to two particle excitations at a given value of $k$ is shown in
Fig.~\ref{fig:J2_0}, computed explicitly by minimizing the sum of energies
of excitations whose pseudomomentum sums to $k$, and similarly for the three
particle excitations. These results are shown by the light and dark shaded
regions in Fig.~\ref{fig:J2_0}. The picture fits well with previously
obtained results.

\begin{figure}
\epsfxsize=\textwidth
\epsfbox{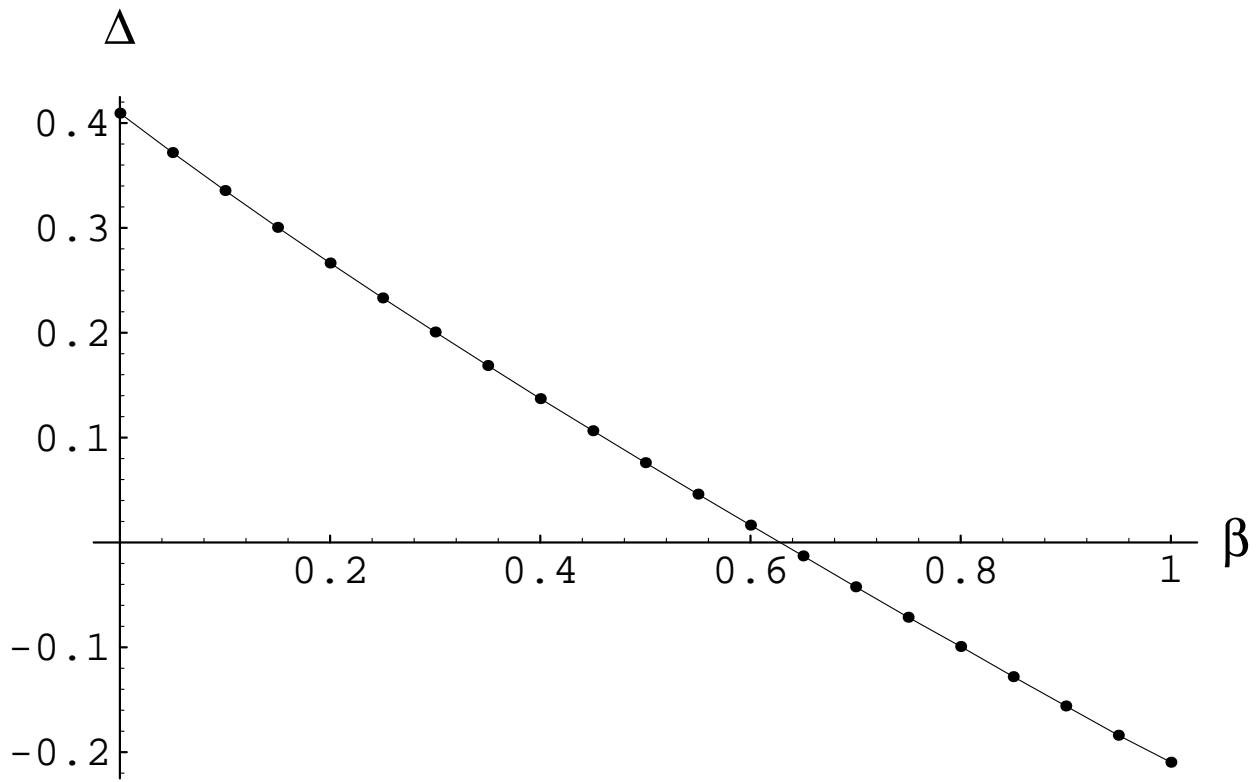}
\caption{The gap $\Delta_\pi$ to the lowest triplet excitation at momentum
	$k=\pi$ as a function of $\beta$ is shown.}
\label{fig:gap}
\end{figure}

We have similarly computed spectra for various values of $\beta$
\cite{TXiang93num,Bursill94den,Schollwock96ons}. The result for the gap to
the lowest lying triplet at $k=\pi$ is shown in Fig.~\ref{fig:gap}. Near
$\beta=0.6$, the excitation spectrum at $k=\pi$ crosses zero and becomes
negative. Our interpretation of this is that our ground state ansatz is
deficient, and this shows up as a condensation of elementary excitations. It
is to be noted that Oitmaa et al. \cite{Oitmaa86cro} also found that
numerically the gap appeared to vanish rapidly near this value of $ \beta $,
although they too were unwilling to conclude that this persisted in the
thermodynamic limit.

Our calculations are consistent with two possible scenarios of what happens
near $ \beta = 0.6 $.  A special value of $ \beta $ could exist where the
gap closes and signals a new phase.  Or, the gap is in fact small and
persists all the way to $ \beta = 1 $ but we do not see it due to our
restricted ansatz for the ground state. Recent DMRG calculations
\cite{Schollwock96ons} have shown to have similar difficulties to estimate
the vanishing gap for $\beta$ close to $1$.  A significant issue appears to
be that the DMRG fixed point seems to invariably lead to a matrix product
ground state that, although it succeeds in reproducing ground state energies
to high accuracy, cannot strictly give a power law decay of spin
correlations.  Thus, we find the ground state energy very accurately at the
Bethe ansatz point $ \beta = 1 $ without finding the expected powerlaw decay
of correlations.  The correlation length spectrum is given by the
eigenvalues \cite{Fannes89exa} of the matrix $\widehat{1}$, and it is hard
to see how this can ever give algebraic correlations.  However, intermediate
correlations for intermediate lengths appear to be well represented in all
cases.

The appendices, App.~\ref{app:expval} - \ref{app:hampole}, contain the
detailed derivations of the results presented in Sec.~\ref{sec:bloch}.
\section{Conclusions}
\label{sec-concl}

The present work suggests that the rapid convergence of the DMRG is
explained by the fact that the states selected are optimally chosen
eigenstates of total block spin. Properly chosen, these states are highly
efficient for building wave functions with a small basis that have low total
spin for all subblocks.

Our analysis also proposes that DMRG inherently predicts exponential decay
of correlations. Nevertheless, fully performed DMRG calculations on systems
with power law decay of correlations seems to agree well with theory. How
this is consistent with our calculations is currently under study.

A related topic is the difficulty to describe the vanishing of the gap close
to a gapless point. However, also ``full'' DMRG calculations seem to suffer
from this problem \cite{Schollwock96ons}.

\appendix

\section{Expectation values in the Bloch states}
\label{app:expval}

In this appendix we will derive expressions for expectation values in the
trial Bloch states $|Q,k)$ of Eq.~(\ref{eq:bloch})
\[
|Q, k )  \equiv \sum_{m=1}^n e^{ikm} { \, \mbox{tr} \left( \;   A[s_n]
\ldots A[s_{m+1}] Q A[s_m] \ldots A[s_1]  \; \right) } | s_n \ldots s_1
\rangle . 
\]
Note that the summation over spins as well as the subscript $n$, the number
of lattice sites, are not explicitly written out.

\subsection{Calculation of the normalization matrix }
We will derive expressions for expectation values of three types of
operators. First we calculate the norm $({{Q}^{\prime}},k | Q,k)$. Then we
show how to obtain the expectation value of total spin, ${\bf S}_T = \sum_i
{\bf S}_i$, where ${\bf S}_i = (S_i^x, S_i^y, S_i^z)$, i.e.\ the expectation
value of the sum of a single site operator. Finally we calculate the
expectation value of the sum of a two site operator like the energy $H =
\sum_i {\bf S}_i \cdot {\bf S}_{i+1}$.  The calculations of these three
types of expectation values differ only in details and not in any
fundamental way. For completeness all three cases are nevertheless covered
in this appendix.

We begin by calculating the norm of the states $|Q,k)$.  Due to the periodic
boundary condition, states with different $k$ are orthogonal.  Using the
definition of $|Q,k)$ we have for the same value of $ k $,
\begin{eqnarray}
({{Q}^{\prime}},k | Q,k ) & = & \sum_m \sum_{{m}^{\prime}} \sum_{ \{ s \} }
e^{-ik{{m}^{\prime}}} e^{ikm} { \, \mbox{tr} \left( \;   A^*[s_n] \ldots
A^*[s_{{{m}^{\prime}}+1}] {{Q}^{\prime}} A[s_{{m}^{\prime}}] \ldots A^*[s_1]
\; \right) } \nonumber \\  
& & \times { \, \mbox{tr} \left( \;   A[s_n] \ldots A[s_{m+1}] Q A[s_m]
\ldots A[s_1]  \; \right) } . \label{eq:qknorm1} 
\end{eqnarray}
We now use the periodic boundary conditions, put ${{m}^{\prime}}=1$ and
change the summation index $m-1 \rightarrow m$.  Using the identities $ { \,
\mbox{tr} \left( \; A \; \right) } { \, \mbox{tr} \left( \; B \; \right) } =
{ \, \mbox{tr} \left( \; A \otimes B \; \right) } $ and $ (A B C) \otimes (D
E F) = (A \otimes D) (B \otimes E) (C \otimes F) $, where the tensor product
is defined in the text after Eq.~(\ref{eq:qopq}), we get
\begin{eqnarray*}
( {{Q}^{\prime}}, k | Q, k ) & = & n \sum_{m=0}^{n-1} e^{ikm} \mbox{tr}
\Bigl( (A^*[s_n] \otimes A[s_n]) (A^*[s_{n-1}] \otimes A[s_{n-1}]) \ldots
(A^*[s_{m+2}] \otimes A[s_{m+2}]) \\
& & \times (\openone \otimes Q) (A^*[s_{m+1}] \otimes A[s_{m+1}]) \ldots
({{Q}^{\prime}}^* \otimes \openone) (A^*[s_1] \otimes A[s_1]) \Bigr) .  
\end{eqnarray*}
By defining 
\begin{eqnarray*}
R_Q & \equiv & \openone \otimes Q \\
L_Q & \equiv & Q^* \otimes \openone
\end{eqnarray*}
and using the definition $\hat{{\sl 1 }} \equiv \sum_s A^*[s] \otimes A[s]$
from Eq.~(\ref{eq:hatdef}) we can rewrite this as
\begin{eqnarray*}
( {{Q}^{\prime}}, k | Q, k ) & = & n \sum_{m=0}^{n-1} e^{ikm} { \, \mbox{tr}
\left( \;   \hat{{\sl 1 }}^{n-m-1} R_Q \hat{{\sl 1 }}^{m} L_{{Q}^{\prime}}
\hat{{\sl 1 }}  \; \right) } \\ 
& = & n \sum_{m=0}^{n} e^{ikm} { \, \mbox{tr} \left( \;   L_{{Q}^{\prime}}
\hat{{\sl 1 }}^{n-m} R_Q \hat{{\sl 1 }}^m  \; \right) } - n 
e^{ikn} { \, \mbox{tr} \left( \;    L_{{Q}^{\prime}} R_Q \hat{{\sl 1 }}^n
\; \right) } , 
\end{eqnarray*}
where we in the last step have added and subtracted the term $m=n$ and used
the cyclicity of the trace. Since $e^{ikn}=1$ we can now write the norm
\begin{eqnarray}
( {{Q}^{\prime}}, k | Q, k ) & = & n \; \sum_{m=0}^{n} { \, \mbox{tr} \left(
\;   L_{{Q}^{\prime}} \hat{{\sl 1 }}^{n-m} R_Q (e^{ik} \hat{{\sl 1 }})^m  \;
\right) } -  { \, \mbox{tr} \left( \;    L_{{Q}^{\prime}} R_Q \hat{{\sl 1
}}^n  \; \right) } \nonumber \\   
& = & n \; { \, \mbox{tr} \left( \;   L_{{Q}^{\prime}} \left( \sum_{m=0}^n
(\hat{{\sl 1 }}^{n-m} R_Q (e^{ik} \hat{{\sl 1 }})^m ) - R_Q \hat{{\sl 1 }}^n
\right)  \; \right) } . \label{eq:qknorm} 
\end{eqnarray}
Let us now introduce the symbol $\Xi$ to represent convolution sums like the
one that appears inside the trace in Eq.~(\ref{eq:qknorm}). Thus, define the
two partition sum $\Xi_n (x,M,y)$ by
\begin{equation}
\Xi_n (x,M,y) \equiv \sum_{m=0}^n x^m M y^{n-m} , \label{eq:twopart}
\end{equation}
where $x, M$ and $y$ are, in our case, square matrices. Later on in this 
section also three partition sums will appear, therefore define 
\begin{equation}
\Xi_n (x,M,y,N,z) \equiv \sum_{m_1=0}^n \sum_{m_2=m_1}^n x^{m_1} M
y^{m_2-m_1} N z^{n-m_2} . \label{eq:threepart}
\end{equation}
Note that the same symbol, $\Xi$, is used to represent both two and three
partition sums; the number of arguments of $\Xi$ determine the number of
summation variables. Using this definition, the norm can now be written as
\begin{equation}
( {{Q}^{\prime}}, k | Q, k ) = n \; { \, \mbox{tr} \left( \;
L_{{Q}^{\prime}} \left( \Xi_n (\hat{{\sl 1 }}, R_Q, (e^{ik} \hat{{\sl 1 }})
) - R_Q \hat{{\sl 1 }}^n \right)  \; \right) } . \label{eq:qknorm2} 
\end{equation}
It is easy to show that $R_Q$ and $L_Q$ commute, so there is no ambiguity in
the order we place the $Q$ and the ${{Q}^{\prime}}$ in terms with
$m={{m}^{\prime}}$ in Eq.~(\ref{eq:qknorm1}).
 
\subsection{Calculation of the total spin}
After finding the norm, we are now interested in the total spin ${\bf S}_T =
\sum_i {\bf S}_i$. We thus need an expression for the expectation value of
the single site operator,
\[
({{Q}^{\prime}},k | ({\bf S}_T)_{op} | Q,k) = \sum_{i=1}^n ({{Q}^{\prime}},k
| ({\bf S}_i)_{op} | Q,k) .
\]
The periodic boundary conditions imply that $( {{Q}^{\prime}}, k | ({\bf
S}_i)_{op} | Q, k )$ is independent of $i$ so let us take $i=1$. We then
have
\begin{eqnarray*}
( {{Q}^{\prime}}, k | ({\bf S}_T)_{op} | Q, k ) & = &  n \sum_{m=1}^n
\sum_{{{m}^{\prime}}=1}^n  e^{-ik{{m}^{\prime}}} 
e^{ikm} { \, \mbox{tr} \left( \; A^*[{{s}^{\prime}}_n] \ldots
A^*[{{s}^{\prime}}_{{{m}^{\prime}}+1}] {{Q}^{\prime}}
A^*[{{s}^{\prime}}_{{m}^{\prime}}] \ldots A^*[{{s}^{\prime}}_1]  \; \right) 
} \\   
& &  \times { \, \mbox{tr} \left( \;   A[s_n] \ldots A[s_{m+1}] Q A[s_m]
\ldots A[s_1]  \; \right) } \langle {{s}^{\prime}}_n \ldots {{s}^{\prime}}_1
| ({\bf S}_1)_{op} | s_n \ldots s_1 \rangle .   
\end{eqnarray*}
To rewrite this expression using $\Xi$ defined in Eq.~(\ref{eq:twopart}) and
(\ref{eq:threepart}) we split the sums over $m$ and ${{m}^{\prime}}$ in
three partial sums
\begin{eqnarray*}
\Sigma_A= &{\displaystyle \sum_{1 \le  m \leq {{m}^{\prime}} } } \\
\Sigma_B= &{\displaystyle \sum_{1 \le  {{m}^{\prime}} \leq m} }\\
\Sigma_F= &{\displaystyle \sum_{1 \le  m = {{m}^{\prime}}} }\\
\end{eqnarray*}
Observing the double counting that appear above we see that
\[
\sum_{m=1}^n \sum_{{{m}^{\prime}}=1}^n = \Sigma_A + \Sigma_B - \Sigma_F . 
\]
Define $S_A, S_B$ and $S_F$ to be the parts of $( {{Q}^{\prime}}, k | ({\bf
S}_T)_{op} | Q, k ) $ with values of $m$ and ${{m}^{\prime}}$ corresponding
to the sums $\Sigma_A, \Sigma_B$ and $\Sigma_F$ respectively.  In a similar
way as for the norm we now get for the sum $A$
\begin{eqnarray}
S_A & = & n \sum_{m=1}^n \sum_{{{m}^{\prime}}=m}^n e^{ik(m-{{m}^{\prime}})}
{ \, \mbox{tr} \left( \;   A^*[s_n] \ldots A^*[s_{{{m}^{\prime}}+1}]
{{Q}^{\prime}}^* \ldots A^*[s_2] A^*[{{s}^{\prime}}_1]  \; \right) } 
\nonumber \\  
& & \times { \, \mbox{tr} \left( \;   A[s_n] \ldots A[s_{m+1}] Q \ldots
A[s_2] A[s_1]  \; \right) } \langle {{s}^{\prime}}_1 | ({\bf S}_1)_{op} |
s_1 \rangle \nonumber \\  
& = & n \sum_{m=1}^n \sum_{{{m}^{\prime}}=m}^n e^{ik(m-{{m}^{\prime}})}
\mbox{tr} \Bigl( (A^*[s_n] \otimes A[s_n]) \ldots (A^*[s_{{{m}^{\prime}}+1}]
\otimes A[s_{{{m}^{\prime}}+1}]) ({{Q}^{\prime}} \otimes \openone )
\nonumber \\
& & \times (A^*[s_{{{m}^{\prime}}}] \otimes A[s_{{{m}^{\prime}}}]) \ldots
(A^*[s_{m+1}] \otimes A[s_{m+1}])(\openone \otimes Q) \nonumber \\
& & \times (A^*[s_{m}] \otimes A[s_{m}]) \ldots (A^*[s_2] \otimes A[s_2])
(A^*[{{s}^{\prime}}_1] \otimes A[s_1]) \Bigr) \langle {{s}^{\prime}}_1 |
({\bf S}_1)_{op} | s_1 \rangle  \nonumber \\ 
& = & n \sum_{m=1}^n \sum_{{{m}^{\prime}}=m}^n e^{ik(m-{{m}^{\prime}})} { \,
\mbox{tr} \left( \;   \hat{{\sl 1 }}^{n-{{m}^{\prime}}} L_{{Q}^{\prime}}
\hat{{\sl 1 }}^{{{m}^{\prime}}-m} R_Q \hat{{\sl 1 }}^{m-1} \hat{S}  \;
\right) } \nonumber \\ 
& = & n \sum_{m=1}^n \sum_{{{m}^{\prime}}=m}^n e^{ik(m-{{m}^{\prime}})} { \,
\mbox{tr} \left( \;   L_{{Q}^{\prime}} \hat{{\sl 1 }}^{{{m}^{\prime}}-m} R_Q
\hat{{\sl 1 }}^{m-1} \hat{S} \hat{{\sl 1 }}^{n-{{m}^{\prime}}}  \; \right) }
, \label{eq:safin} 
\end{eqnarray}
where we have used the definition of $\hat{S}$ from Eq.~(\ref{eq:hatdef}). 
By changing summation index $m-1 \rightarrow m$ and ${{m}^{\prime}}-1
\rightarrow {{m}^{\prime}}$ we get
\begin{eqnarray*}
S_A & = & n \sum_{m=0}^{n-1} \sum_{{{m}^{\prime}}=m}^{n-1}
e^{ik(m-{{m}^{\prime}})} { \, \mbox{tr} \left( \; L_{{Q}^{\prime}} \hat{{\sl
1 }}^{{{m}^{\prime}}-m} R_Q \hat{{\sl 1 }}^{m} \hat{S} \hat{{\sl 1
}}^{n-1-{{m}^{\prime}}}  \; \right) } \\ 
& = & n \, \mbox{tr} \left( \;   L_{{Q}^{\prime}} \Xi_{n-1}( e^{-ik}
\hat{{\sl 1 }}, R_Q, \hat{{\sl 1 }}, \hat{S}, \hat{{\sl 1 }} )  \; \right) .
\end{eqnarray*}
In a similar way we get for the sum $B$ 
\begin{eqnarray}
S_B & = & n \sum_{{{m}^{\prime}}=1}^n \sum_{m={{m}^{\prime}}}^n
e^{ik(m-{{m}^{\prime}})}  { \, \mbox{tr} \left( \; A[s_n] \ldots
A[s_{{{m}^{\prime}}+1}] {{Q}^{\prime}} \ldots A[s_2] A[{{s}^{\prime}}_1]  \;
\right) }^* \nonumber \\ 
& & \times { \, \mbox{tr} \left( \;   A[s_n] \ldots A[s_{m+1}] Q \ldots
A[s_2] A[s_1]  \; \right) } \langle {{s}^{\prime}}_1 | ({\bf S}_1)_{op} | s_1
\rangle \label{eq:sb} \\  
& = & n { \, \mbox{tr} \left( \;   L_Q \Xi_{n-1} \left( \hat{{\sl 1 }},
\hat{S}, \hat{{\sl 1 }}, R_Q, e^{ik} \hat{{\sl 1 }} \right)  \; \right) }
. \label{eq:sbfin} 
\end{eqnarray}
It is also possible to show that
\begin{displaymath}
S_B({{Q}^{\prime}},Q) = \left( S_A(Q,{{Q}^{\prime}}) \right)^* .
\end{displaymath}
The sum $F$ contains the terms that are counted twice in $A$ and $B$ and
$S_F$ should therefore be subtracted from $S_A+S_B$. We get
\begin{eqnarray}
S_F & = & n \sum_{m=1}^n \; { \, \mbox{tr} \left( \;   \hat{{\sl 1 }}^{n-m}
L_{{Q}^{\prime}} R_Q \hat{{\sl 1 }}^{m-1} \hat{S}  \; \right) } \nonumber \\
& = & n \sum_{m=0}^{n-1} { \, \mbox{tr} \left( \;   L_{{Q}^{\prime}} R_Q
\hat{{\sl 1 }}^{m} \hat{S} \hat{{\sl 1 }}^{n-1-m}  \; \right) } \nonumber \\
& = & n \; { \, \mbox{tr} \left( \;   L_{{Q}^{\prime}} R_Q \Xi_{n-1}(
\hat{{\sl 1 }}, \hat{S}, \hat{{\sl 1 }})  \; \right) } . \label{eq:sgfin}  
\end{eqnarray}
We now collect the results from Eq.~(\ref{eq:safin}), (\ref{eq:sbfin}) and
(\ref{eq:sgfin}). The expectation value of the total spin is thus
\begin{eqnarray}
( {{Q}^{\prime}}, k | (S_T)_{op} | Q, k ) & = & S_A({{Q}^{\prime}},Q) +
S_B({{Q}^{\prime}},Q) - S_F({{Q}^{\prime}},Q) \nonumber \\
& = & n \; \mbox{tr} \left( \right. L_{{Q}^{\prime}} \times \nonumber \\  
& &  \left[ \right. \Xi_{n-1} ( e^{-ik} \hat{{\sl 1 }}, R_Q, \hat{{\sl 1 }},
\hat{S}, \hat{{\sl 1 }} ) \nonumber \\  
& &  \mbox{} + \Xi_{n-1} ( \hat{{\sl 1 }}, \hat{S} , \hat{{\sl 1 }} ,R_Q ,
e^{ik} \hat{{\sl 1 }} )  \nonumber \\     
& & \mbox{} - R_Q \; \Xi_{n-1} ( \hat{{\sl 1 }}, \hat{S}, \hat{{\sl 1 }} ) 
\left. \right] \left. \right) . \label{eq:qkstot}  
\end{eqnarray}
We have here not made use of the fact that $S_B$ can be determined from
${S_A}^*$.

\subsection{Calculation of the energy}
The final operator we need is the energy $H = \sum_i h_{i,i+1}$, where
$h_{i,i+1} = {\bf S}_i \cdot {\bf S}_{i+1}$.  We thus have to find an
expression for the expectation value of a two site operator. The procedure
to find it is analogous to how we found the total spin. We use the periodic
boundary conditions to put $i=1$. Thus
\begin{eqnarray*}
( {{Q}^{\prime}}, k |H_{op}| Q, k ) & = & \sum_{i=1}^n \sum_{m=1}^n
\sum_{{{m}^{\prime}}=1}^n 
e^{-ik{{m}^{\prime}}} e^{ikm} { \, \mbox{tr} \left( \; A[s_n] \ldots
A[s_{m+1}] Q A[s_m] \ldots A[s_1]  \; \right) } \\  
& & \times { \, \mbox{tr} \left( \;   A^*[{{s}^{\prime}}_n] \ldots 
A^*[{{s}^{\prime}}_{{{m}^{\prime}}+1}] {{Q}^{\prime}}
A^*[{{s}^{\prime}}_{{m}^{\prime}}] \ldots A^*[{{s}^{\prime}}_1]  \; \right)
} \langle {{s}^{\prime}}_n \ldots {{s}^{\prime}}_1 | h_{1,2} | s_n \ldots
s_1 \rangle .   
\end{eqnarray*}
Since the terms with $m=1$ and/or ${{m}^{\prime}}=1$ in the expression above
are special in the sense that the matrices $Q$ and ${{Q}^{\prime}}$ mix with
the operator $h_{1,2}$, we this time have to split the sum in six partial
sums
\begin{eqnarray*}
\Sigma_A= &{\displaystyle \sum_{2 \le  m \leq {{m}^{\prime}} } } \\
\Sigma_B= &{\displaystyle \sum_{ 2 \le  {{m}^{\prime}} \leq m} }\\
\Sigma_F= &{\displaystyle \sum_{ 2 \le m = {{m}^{\prime}}} }\\
\Sigma_C= &{\displaystyle \sum_{ m = 1 < {{m}^{\prime}} } }\\
\Sigma_D= &{\displaystyle \sum_{ {{m}^{\prime}} = 1 < m } }\\
\Sigma_E= &{\displaystyle  \sum_{ m = {{m}^{\prime}} = 1 }}
\end{eqnarray*}
We note that
\[
\sum_{m=1}^n \sum_{{{m}^{\prime}}=1}^n = \Sigma_A  + \Sigma_B - \Sigma_F +
\Sigma_C + \Sigma_D + \Sigma_E .
\]
Analogous to what was done for the total spin, we define $H_A, H_B$ etc.\ to
be the parts of $( {{Q}^{\prime}}, k |H_{op}| Q, k )$ with values of $m$ and
${{m}^{\prime}}$ corresponding to $\Sigma_A, \Sigma_B$ etc. The sum $H_A$
for the two site operator is very similar to $S_A$ in Eq.~(\ref{eq:safin})
for the single particle operator. We find
\begin{eqnarray}
H_A & = & n \sum_{m=2}^n \sum_{{{m}^{\prime}}=m}^n e^{ik(m-{{m}^{\prime}})}
\mbox{tr} \Bigl( (A^*[s_n] \otimes A[s_n]) (A^*[s_{n-1}] \otimes
A[s_{n-1}]) \ldots (A^*[s_{{{m}^{\prime}}+1}] \otimes
A[s_{{{m}^{\prime}}+1}]) \nonumber \\
& & \times ({{Q}^{\prime}} \otimes \openone )(A^*[s_{{{m}^{\prime}}}]
\otimes A[s_{{{m}^{\prime}}}]) \ldots (A^*[s_{m+1}] \otimes A[s_{m+1}])(
\openone \otimes Q) \nonumber \\
& & \times (A^*[s_{m}] \otimes A[s_{m}]) \ldots (A^*[{{s}^{\prime}}_2]
\otimes A[s_2]) (A^*[{{s}^{\prime}}_1] \otimes A[s_1]) \Bigr) \langle
{{s}^{\prime}}_2, {{s}^{\prime}}_1 | h_{1,2} | s_2, s_1 \rangle \nonumber \\
& = & n \sum_{m=2}^n \sum_{{{m}^{\prime}}=m}^n e^{ik(m-{{m}^{\prime}})} { \,
\mbox{tr} \left( \;   L_{{Q}^{\prime}} \hat{{\sl 1 }}^{{{m}^{\prime}}-m} R_Q
\hat{{\sl 1 }}^{m-2} \hat{S} \hat{S} \hat{{\sl 1 }}^{n-{{m}^{\prime}}}  \;
\right) } , \label{eq:hafin} 
\end{eqnarray}
where we have used the hat mapping defined in Eq.~(\ref{eq:hatdef}) for the
Hamiltonian matrix $\hat{S} \hat{S}$. By changing summation indices $m-2
\rightarrow m$ and ${{m}^{\prime}}-2 \rightarrow {{m}^{\prime}}$, and using
the $\Xi$ notation for the sum, we get
\begin{eqnarray*}
H_A & = & n \sum_{m=0}^{n-2} \sum_{{{m}^{\prime}}=m}^{n-2}
e^{ik(m-{{m}^{\prime}})} { \, \mbox{tr} \left( \; L_{{Q}^{\prime}} \hat{{\sl
1 }}^{{{m}^{\prime}}-m} R_Q \hat{{\sl 1 }}^{m} \hat{S} \hat{S} \hat{{\sl 1
}}^{n-2-{{m}^{\prime}}}  \; \right) } \\ 
& = & n { \, \mbox{tr} \left( \;   L_{{Q}^{\prime}} \Xi_{n-2}( e^{-ik}
\hat{{\sl 1 }}, R_Q, \hat{{\sl 1 }}, \hat{S} \hat{S}, \hat{{\sl 1 }} )  \;
\right) } .  
\end{eqnarray*}
In a similar way we get for the sum $H_B$ 
\begin{eqnarray}
H_B & = & n \sum_{{{m}^{\prime}}=2}^n \sum_{m={{m}^{\prime}}}^n
e^{ik(m-{{m}^{\prime}})}  { \, \mbox{tr} \left( \;   A[s_n] \ldots
A[s_{{{m}^{\prime}}+1}] {{Q}^{\prime}} \ldots A[{{s}^{\prime}}_2]
A[{{s}^{\prime}}_1]  \; \right) }^* \nonumber \\
& & \times { \, \mbox{tr} \left( \;   A[s_n] \ldots A[s_{m+1}] Q \ldots
A[s_2] A[s_1]  \; \right) } \langle {{s}^{\prime}}_2, {{s}^{\prime}}_1 |
h_{1,2} | s_2, s_1 \rangle \label{eq:hb} \\ 
& = & n { \, \mbox{tr} \left( \;   L_Q \Xi_{n-2} \left( \hat{{\sl 1 }},
\hat{S} \hat{S}, \hat{{\sl 1 }}, R_Q, e^{ik} \hat{{\sl 1 }} \right)  \;
\right) } . \label{eq:hbfin} 
\end{eqnarray}
It is also possible to show that
\begin{equation}
H_B({{Q}^{\prime}},Q) = \left( H_A(Q,{{Q}^{\prime}}) \right)^*
. \label{eq:hastar} 
\end{equation}
The sum $F$ contains the terms that are counted twice in $A$ and $B$ and
$H_F$ should therefore be subtracted from $H_A+H_B$. In the same way as we
found $S_F$ in Eq.~(\ref{eq:sgfin}) we now find
\begin{eqnarray}
H_F & = & n \sum_{m=2}^n \; { \, \mbox{tr} \left( \;   \hat{{\sl 1 }}^{n-m}
L_{{Q}^{\prime}} R_Q \hat{{\sl 1 }}^{m-2} \hat{S} \hat{S}  \; \right) }
\nonumber \\ 
& = & n \; { \, \mbox{tr} \left( \;   L_{{Q}^{\prime}} R_Q \Xi_{n-2}(
\hat{{\sl 1 }}, \hat{S} \hat{S}, \hat{{\sl 1 }})  \; \right) }
. \label{eq:hgfin}  
\end{eqnarray}
The sums $C,D$ and $E$ contain terms were the matrix $Q$ and/or
${{Q}^{\prime}}$ mixes with the operator $h_{1,2}$. For $C$ we get
\begin{eqnarray}
H_C & = & n \sum_{{{m}^{\prime}}=2}^n e^{-ik({{m}^{\prime}}-1)} \; { \,
\mbox{tr} \left( \;   A^*[s_n] \ldots A^*[s_{{{m}^{\prime}}+1}]
{{Q}^{\prime}}^* \ldots A^*[{{s}^{\prime}}_2] A^*[{{s}^{\prime}}_1]  \;
\right) } \nonumber \\
& & \times { \, \mbox{tr} \left( \;   A[s_n] \ldots A[s_2] Q A[s_1]  \;
\right) } \langle {{s}^{\prime}}_2, {{s}^{\prime}}_1 | h_{1,2} | s_2, s_1
\rangle \label{eq:hc} \\ 
& = &  n \sum_{{{m}^{\prime}}=2}^n e^{-ik({{m}^{\prime}}-1)} { \, \mbox{tr}
\left( \;   \hat{{\sl 1 }}^{n-{{m}^{\prime}}} L_{{Q}^{\prime}} 
\hat{{\sl 1 }}^{{{m}^{\prime}}-2} \hat{S} R_Q \hat{S}  \; \right) }
\nonumber \\ 
& = & n e^{-ik} { \, \mbox{tr} \left( \;   L_{{Q}^{\prime}} \Xi_{n-2}(
e^{-ik} \hat{{\sl 1 }}, \hat{S} R_Q \hat{S}, \hat{{\sl 1 }} )  \; \right) }
. \label{eq:hcfin} 
\end{eqnarray}
$H_D$ yields
\begin{eqnarray}
H_D & = & n \sum_{m=2}^n e^{ik(m-1)} { \, \mbox{tr} \left( \;   A^*[s_n]
\ldots A^*[{{s}^{\prime}}_2] {{Q}^{\prime}} A^*[{{s}^{\prime}}_1]  \;
\right) } \nonumber \\ 
& & \times { \, \mbox{tr} \left( \;   A[s_n] \ldots A[s_{m+1}] Q \ldots
A[s_2] A[s_1]  \; \right) } \langle {{s}^{\prime}}_2, {{s}^{\prime}}_1 |
h_{1,2} | s_2, s_1 \rangle . \label{eq:hd} \\ 
& = & n \sum_{m=2}^n e^{ik(m-1)} { \, \mbox{tr} \left( \;   \hat{{\sl 1
}}^{n-m} R_Q \hat{{\sl 1 }}^{m-2} \hat{S} L_{{Q}^{\prime}} \hat{S}  \;
\right) } \nonumber \\ 
& = & n e^{-ik} { \, \mbox{tr} \left( \;   L_{{Q}^{\prime}} \hat{S} \;
\Xi_{n-2} \left( e^{-ik} \hat{{\sl 1 }}, R_Q, \hat{{\sl 1 }}  \right)
\hat{S}  \; \right) } , \label{eq:hdfin} 
\end{eqnarray}
where we in the last step used that $e^{-ikn}=1$. One can also show that
\begin{displaymath}
H_D({{Q}^{\prime}},Q) = (H_C(Q,{{Q}^{\prime}}))^* .
\end{displaymath}
The ``sum'' $E$ is just
\begin{eqnarray}
H_E & = & n \; { \, \mbox{tr} \left( \;   \hat{{\sl 1 }}^{n-2} \hat{S} R_Q
L_{{Q}^{\prime}} \hat{S}  \; \right) } \nonumber \\ 
& = & n \; { \, \mbox{tr} \left( \;   L_{{Q}^{\prime}} \hat{S} \hat{{\sl 1
}}^{n-2} \hat{S} R_Q  \; \right) } . \label{eq:hefin} 
\end{eqnarray}
We now collect the results from Eq.~(\ref{eq:hafin}), (\ref{eq:hbfin}),
(\ref{eq:hgfin}), (\ref{eq:hcfin}), (\ref{eq:hdfin}) and
(\ref{eq:hefin}). For the whole Hamiltonian we thus have
\begin{eqnarray}
( {{Q}^{\prime}}, k |H| Q, k ) & = & H_A({{Q}^{\prime}},Q) +
H_B({{Q}^{\prime}},Q) - H_F({{Q}^{\prime}},Q) \nonumber \\ 
& & \mbox{} + H_C({{Q}^{\prime}},Q) + H_D({{Q}^{\prime}},Q) +
H_E({{Q}^{\prime}},Q) \\ 
& = & n \; \mbox{tr} \left( \right. L_{{Q}^{\prime}} \times \nonumber \\  
& &  \left[ \right. \Xi_{n-2} ( e^{-ik} \hat{{\sl 1 }}, R_Q, \hat{{\sl
1 }}, \hat{S} \hat{S}, \hat{{\sl 1 }} ) \nonumber \\  
& &  \mbox{} + \Xi_{n-2} ( \hat{{\sl 1 }}, \hat{S} \hat{S}, \hat{{\sl 1 }},
R_Q , e^{ik} \hat{{\sl 1 }} ) \nonumber \\    
& & \mbox{} - R_Q \; \Xi_{n-2} ( \hat{{\sl 1 }}, \hat{S} \hat{S}, \hat{{\sl 1
}} ) \nonumber \\   
& & \mbox{} + e^{-ik} \Xi_{n-2} ( e^{-ik} \hat{{\sl 1 }}, \hat{S} R_Q \hat{S},
\hat{{\sl 1 }} ) \nonumber \\
& & \mbox{} + e^{-ik} \hat{S} ( \Xi_{n-2} (e^{-ik} \hat{{\sl 1 }}, R_Q,
\hat{{\sl 1 }} ) ) \hat{S} \nonumber \\ 
& & \mbox{} + \hat{S} \hat{{\sl 1 }}^{n-2} \hat{S} R_Q \left. \right]
\left. \right) . \label{eq:qkham} 
\end{eqnarray}
We have here not made use of the relations
$H_B({{Q}^{\prime}},Q)=(H_A(Q,{{Q}^{\prime}}))^*$ and
$H_D({{Q}^{\prime}},Q)=(H_C(Q,{{Q}^{\prime}}))^*$.   

Eq.~(\ref{eq:qknorm}), (\ref{eq:qkstot}) and (\ref{eq:qkham}) now contain
the desired expectation values, expressed in terms of convolution
sums. These sums can be expediently calculated using recursive relations, as
we will show in the next section.

\section{Calculating the partition sums recursively}
\label{app:recsum}

Expectation values between the Bloch states $|Q,k)$ can be divided into
partial sums with the general forms of two-partition and three-partition
sums defined in Eq.~(\ref{eq:twopart}) - (\ref{eq:threepart}).  The number
of terms in the two-partition sum with upper limit $n$ is $n+1$ while the
number of terms in the three-partition sum with upper limit $n$ is
$\frac{(n+1)(n+2)}{2}$.  Both these sums can be calculated recursively with
a number of operations of the order $\log(n)$. For the two-partition sum
Eq.~(\ref{eq:twopart}) we find that the sum with upper limit $2 n$ can be
found from the sum with upper limit $n$ by
\begin{displaymath}
\left\{
\begin{array}{rcl}
\Xi_{2n} (x,S,y) & = & x^n \; \Xi_n (x,S,y) + \Xi_n (x,S,y) \; y^n - x^n S 
y^n \\ 
x^{2n} & = & x^n x^n \\
y^{2n} & = & y^n y^n
\end{array}
\right.
\end{displaymath}
with the starting sum
\begin{displaymath}
\Xi_1 (x,S,y) = x S + S y .
\end{displaymath}
We thus get sums where $n = 2^j$, $j$ integer.  Each recursion step requires
a constant number of additions and multiplications which implies a total
computational effort of order $\log(n)$.  The three-partition sum,
Eq.~(\ref{eq:threepart}), can be done in a similar way. Here the $2 n - 2$
sum is reached from the $n-2$ sum by
\begin{displaymath}
\left\{
\begin{array}{rcl}
\Xi_{2n-2} (x,S,y,T,z) & = & x^n \; \Xi_{n-2} (x,S,y,T,z) + \Xi_{n-2}
(x,S,y,T,z) \; z^n \\
& & \mbox{} + \Xi_{n-1} (x,S,y) \; \Xi_{n-1} (y,T,z) \\
\Xi_{2n-1} (x,S,y) & = & x^n \; \Xi_{n-1} (x,S,y) + \Xi_{n-1} (x,S,y) \; y^n
\\ 
x^{2n} & = & x^n x^n 
\end{array}
\right.
\end{displaymath}
with similar expressions for $\Xi_{2n-1} (y,T,z), \; y^{2n}$ and $z^{2n}$.  
Here we start with
\begin{displaymath}
\left\{
\begin{array}{rcl}
\Xi_0 (x,S,y,T,z) & = & ST \\
\Xi_1 (x,S,y) & = & x S + S y \\
\Xi_1 (y,T,z) & = & y T + T z 
\end{array}
\right.
\end{displaymath}
and we get sums with upper summation bound $n-2$, with $n=2^j$ and $j$ an
integer. Also here the computational effort is of order $\log(n)$.  In this
recursion scheme we also get the two-partition sum with upper bound $n-1$.

\section{The Pole Expansion}
\label{app:pole}
Although calculating the sums recursively is a nice method for finite size
chains, we would like to calculate the expectation values in the limit $n
\rightarrow \infty$. As we will show in this section, it is actually
possible to do this directly by analyzing the sums' asymptotic form.  In the
next section, App.~\ref{app:hampole}, we apply the results to the actual
sums in the expectation values of App.~\ref{app:expval}.

\subsection{Three-partition sums}
In App.~\ref{app:expval} expectation values were calculated and expressed in
terms of sums. These sums are of the general form
\[
S_n = \sum_{n_1,n_2,n_3 \geq 0} (\gamma x)^{n_1} S x^{n_2} T x^{n_3} 
\delta_{n,n_1+n_2+n_3} ,
\]
where $x$, $S$ and $T$ are $m^2 \times m^2$ matrices and $\gamma = e^{ik}$
is a phase factor. We would like to know the asymptotic form of $S_n$ as $n
\rightarrow \infty$. This form can be found if we take the $z$-transform
(also known as discrete Laplace transform) of $S_n$ and then analyze the
pole structure of the transformed sum. Define the $z$-transform of the sum
$S_n$ by
\[
F^S[\lambda] \equiv \sum_{n=0}^\infty \lambda^n S_n .
\]
We then have 
\begin{eqnarray*}
F^S[\lambda] & = & \sum_{n=0}^\infty \sum_{n_1,n_2,n_3} (\lambda \gamma
x)^{n_1} S (\lambda x)^{n_2} T (\lambda x)^{n_3} \delta_{n,n_1+n_2+n_3} \\ 
& = & \sum_{n_1,n_2,n_3} (\lambda \gamma x)^{n_1} S (\lambda x)^{n_2} T
(\lambda x)^{n_3} \\
& = & ( \sum_{n_1=0}^\infty (\lambda \gamma x)^{n_1} ) S (
\sum_{n_2=0}^\infty (\lambda x )^{n_2} ) T ( \sum_{n_3=0}^\infty (\lambda
x)^{n_3} ) .
\end{eqnarray*}
Let us define $U$ as the matrix that diagonalizes $x$. Let us also define
a transformation $M^D$ of a general $m^2 \times m^2$ matrix $M$ by
\[
M^D = U^{-1} M U .
\]
Thus ${x^D}$ is a diagonal matrix with the eigenvalues of $x$ on the
diagonal, while the transformation $M^D$ of a general matrix $M$ need not be
diagonal.  We then have
\begin{eqnarray*}
F^S[\lambda] & = & U \: \left( \sum_{n_1=0}^\infty (\lambda \gamma
{x^D})^{n_1} \right) \: U^{-1} S U \: \left( \sum_{n_2=0}^\infty (\lambda
{x^D})^{n_2} \right) \: U^{-1} T U \: \left( \sum_{n_3=0}^\infty (\lambda
{x^D})^{n_3} \right) \: U^{-1} \nonumber \\ 
& = & U \left( \begin{array}{cc} \frac{1}{1-\lambda \gamma x_1} & 0 \\ 
				  0 &  \ddots 
		\end{array} \right) 
{S^D} \left( \begin{array}{ccc} 	\frac{1}{1-\lambda x_1} & & \\
					& & \ddots 
		\end{array} \right) 
{T^D} \left( \begin{array}{cc} 	\frac{1}{1-\lambda x_1} &  \\
					&  \ddots 
		\end{array} \right) U^{-1} ,
\end{eqnarray*}
where $x_i$ are the eigenvalues of $x$.  In our case ${x^D}$ are the
diagonalized $\hat{{\sl 1 }}$ and $x_i$ are eigenvalues of $\hat{{\sl 1
}}$. The largest eigenvalue of $\hat{{\sl 1 }}$ is $ x_1 = 1$ and the other
eigenvalues have absolute values less than $0.8$.  The order of the poles of
$F^S[\lambda]$ will be different for $k=0$ and $k \neq 0$. We will therefore
have to treat these two cases separately.  We first determine the asymptotic
form in the $k=0$ case.

\subsubsection{Pole expansion for zero momentum}
The transform will now have as elements 
\begin{equation}
(F^S[\lambda])^{i,j} = \sum_l ({S^D})^{i,l} ({T^D})^{l,j} \frac{1}{(1 -
\lambda x_i)} \frac{1}{(1 - \lambda y_l)} \frac{1}{(1 - \lambda z_j)} . 
\label{eq:Flam0}  
\end{equation}
Note that we have for simplicity not written out the leading $U$ and the
trailing $U^{-1}$ in the above formula. Also in the rest of this article,
these $U$ and $U^{-1}$ will be omitted.  Since the largest eigenvalue of
$\hat{{\sl 1 }}$ is $x_1 = 1$ and the next highest is $x_2 \approx 0.8$, we
take as an ansatz for the behavior of $S_n$ for large $n$
\begin{displaymath}
S_n = A n^2 + B n + C + \; \mbox{corrections} ,
\end{displaymath}
where the corrections are of order $x_2^n \approx 0.8^n $ and thus very
small. We now calculate $F^S[\lambda]$ using this asymptotic form of
$S_n$. Call it $F^A[\lambda]$ to distinguish it from the original form.
\begin{eqnarray} 
F^A[\lambda] & = & \sum_{n=0}^\infty \lambda^n (A n^2 + B n + C) \nonumber
\\ 
& = &   A \sum n^2 \lambda^n + B \sum n \lambda^n + C \sum \lambda^n
\nonumber \\
& = & A \frac{2 \lambda^2}{(1-\lambda)^3} + (A+B)
\frac{\lambda}{(1-\lambda)^2} + C \frac{1}{(1-\lambda)} \nonumber \\  
& = & \frac{2A}{(1-\lambda)^3} + \frac{B-3A}{(1-\lambda)^2} +
\frac{C-B+A}{(1-\lambda)} . \label{eq:polex}
\end{eqnarray}
We see that $F^A[\lambda]$ in Eq.~(\ref{eq:polex}) has poles at
$\lambda=1$. $F^S[\lambda]$ also has poles at $\lambda=1$ and is analytical
in a neighborhood.  We therefore expand $F^S[\lambda]$ around $\lambda=1$
and identify terms. This will also justify the asymptotic form we have
suggested above.  Noting that $ x_1 = 1 $ we define a function $g(\lambda)$
by
\begin{eqnarray*}
g ( \lambda )   & \equiv &   ( 1 - \lambda ) ( 1 - \lambda {x^D} )^{-1}  \\
	 & = &   \left( \begin{array}{ccc} 	1 & & \\
					& \frac{1 - \lambda }{1-\lambda x_2} & \\
					& & \ddots 
		\end{array} \right) .
\end{eqnarray*}
We then note that
\begin{equation}
(1 - \lambda )^3  F^S[\lambda] =  g( \lambda ) {S^D} g( \lambda ) {T^D}
g ( \lambda ) . \label{eq:fs}
\end{equation}
We use the shorthand notation $ g \equiv g ( 1 ) $ , $ {{g}^{\prime}} \equiv
{{g}^{\prime}} ( 1 ) $ and $ {g^{\prime \prime } } \equiv {g^{\prime \prime
} } ( 1 ) $. Combining Eq.~(\ref{eq:polex}) and (\ref{eq:fs}) we arrive at
the central result of the pole expansion for the three-partition sum when $k
= 0$:
\begin{equation}
\left\{
\begin{array}{rcl}
2 A  & = &  \lim_{ \lambda \rightarrow 1 } \left(  ( 1 - \lambda )^3
   F^S[\lambda ] \right) \nonumber \\
& = &  g \tilde{S} g \tilde{T} g  \\
-(B - 3 A)  & = & \lim_{ \lambda \rightarrow 1 } \left( \frac{d}{d \lambda }
   (1 - \lambda )^3 F^S[\lambda ]  \right) \nonumber \\ 
&  = &   {{g}^{\prime}}  {S^D} g   {T^D} g   + g   {S^D} {{g}^{\prime}}
   {T^D} g   + g   {S^D} g   {T^D} {{g}^{\prime}}  \\ 
2(C - B + A) & = & \lim_{ \lambda \rightarrow 1 } \left( \frac{d^2}{d 
   \lambda^2} ( 1 - \lambda )^3 F^S[\lambda ] \right) \nonumber \\  
& = & {g^{\prime \prime } }  {S^D} g   {T^D} g  + g  {S^D} {g^{\prime \prime
   } }   {T^D}  g  + g  {S^D} g {T^D} {g^{\prime \prime } }  \nonumber \\ 
&   &  \mbox{} + 2 g {S^D} {{g}^{\prime}} {T^D} {{g}^{\prime}} + 2
   {{g}^{\prime}} {S^D} g {T^D} {{g}^{\prime}} + 2 {{g}^{\prime}}  {S^D}
   {{g}^{\prime}} {T^D} g . 
\end{array} \label{eq:3pole0}
\right.
\end{equation}

We note that 
\begin{eqnarray*}
{{g}^{\prime}}(\lambda) & = & \left( \begin{array}{ccc} 0 & & \\ & 
\frac{-(1-x_2)}{(1-\lambda x_2)^2} & \\ & & \ddots \end{array} \right) \\  
{g^{\prime \prime } }(\lambda) & = & \left( \begin{array}{ccc} 0 & & \\ &
\frac{-2 x_2 (1-x_2)}{(1-\lambda x_2)^3} & \\ & & \ddots \end{array} \right)
. \\ 
\end{eqnarray*}

\subsubsection{Pole expansion for nonzero momentum}
We now treat the case when the crystal momentum $k \neq 0$.  In this case
the first matrix $x$ is multiplied by a phasefactor $\gamma = e^{ik} \neq 1$
and we have the elements
\begin{equation}
(F^S[\lambda])^{i,j} = \sum_l ({S^D})^{i,l} ({T^D})^{l,j} \frac{1}{(1 -
\lambda \gamma x_i)} \frac{1}{(1 - \lambda y_l)} \frac{1}{(1 - \lambda z_j)}
. \label{eq:Flam1} 
\end{equation}
We notice that this time there can be no poles of order three at $\lambda =
1$. Instead we have a pole at $\lambda = \gamma^{-1}$. The asymptotic form
now looks like
\begin{equation}
S_n = B n + C + \gamma^n {{C}^{\prime}} + \; \mbox{corrections}
. \label{eq:3asymk} 
\end{equation}
The new term ${{C}^{\prime}}$ will give rise to a term $(\lambda -
\gamma^{-1})^{-1}$ and to match this term we have to expand around $\lambda
= \gamma^{-1}$. There can only be a simple pole at $\lambda = \gamma^{-1}$
so there will not be any terms ${{A}^{\prime}}$ or ${{B}^{\prime}}$ (i.e.\
terms proportional to $\gamma^n n^2$ or $\gamma^n n$). We have
\begin{eqnarray}
F^A[\lambda] & = & \sum_{n=0}^\infty \lambda^n (B n + C + \gamma^n
{{C}^{\prime}}) \nonumber \\
& = & \frac{B}{(\lambda-1)^2} + \frac{B-C}{(\lambda-1)} +
\frac{-{{C}^{\prime}}}{(\lambda \gamma - 1)} \nonumber \\
& = & \frac{B}{(1-\lambda)^2} + \frac{C-B}{(1-\lambda)} +
\frac{{{C}^{\prime}}}{(1 - \lambda \gamma)} . \label{eq:fak}
\end{eqnarray}
By defining a function $h(\lambda)$ 
\begin{eqnarray*}
h (\lambda) & \equiv & (1-\lambda {x^D})^{-1} \\
& = & \left( \begin{array}{ccc} \frac{1}{(1-\lambda x_1)} & & \\ & 
\frac{1}{(1-\lambda x_2)} & \\ & & \ddots \end{array} \right) , 
\end{eqnarray*}
and using the definition of $g(\lambda)$ we write $F^S[\lambda]$ in the
following two ways
\begin{eqnarray*}
F^S[\lambda] & = & \frac{1}{(1-\lambda)^2} h (\lambda \gamma) {S^D} g
(\lambda) {T^D} g (\lambda) \\
& = & \frac{1}{(1-\lambda \gamma)} g (\lambda \gamma) {S^D} h (\lambda) 
{T^D} h (\lambda) .
\end{eqnarray*}
In a similar manner to the $k=0$ case we now find 
\begin{equation}
\left\{
\begin{array}{rcl}
B & = & \lim_{ \lambda \rightarrow 1} \left( (1-\lambda)^2 F^S[\lambda]
\right) \\
& = & h(\gamma) {S^D} g {T^D} g  \\
-(C-B) & = & \lim_{ \lambda \rightarrow 1} \left( \frac{d}{d \lambda}
(1-\lambda)^2 F^S[\lambda] \right) \\
& = & \gamma {{ h }^{\prime}} (\gamma) {S^D} g {T^D} g + h (\gamma) {S^D}
{{g}^{\prime}} {T^D} g + h (\gamma) {S^D} g {T^D} {{g}^{\prime}} \\ 
{{C}^{\prime}} & = & \lim_{ \lambda \rightarrow \gamma^{-1}} \left(
(1-\lambda \gamma) F^S[\lambda] \right)  \\ 
& = & g {S^D} h (\gamma^{-1}) {T^D} h (\gamma^{-1}) .
\end{array}
\right.
\label{eq:3polek}
\end{equation}

\subsection{Two-partition sums}
The pole expansion can of course also be done for the two-partition sums
defined in Eq.~(\ref{eq:twopart}). We will not go through the details since
the calculation is analogous to the three-partition case but for
completeness only list the results.

Let us analyze the sum
\begin{equation}
S_n = \sum_{m=0}^n (\gamma x)^m S x^{n-m} ,
\end{equation}
where $\gamma$,$x$ and $S$ are defined as before. For the case
$\gamma=e^{ik}=1$
the asymptotic form as $n \rightarrow \infty$ is
\begin{equation}
S_n = B n + C + \mbox{corrections} , \label{eq:twoasym0}
\end{equation}
with
\begin{eqnarray}
\left\{
\begin{array}{rcl}
B & = & g {S^D} g \\
-(C-B) & = & {{g}^{\prime}} {S^D} g + g {S^D} {{g}^{\prime}} .
\end{array}
\right. \label{eq:twopole0}
\end{eqnarray}
For the case $\gamma \neq 1$ we instead get the asymptotic form
\begin{equation}
S_n = C + \gamma^n {{C}^{\prime}} + \mbox{corrections} , \label{eq:twoasymk}
\end{equation}
with
\begin{equation}
\left\{
\begin{array}{rcl}
C & = & h (\gamma) {S^D} g \\
{{C}^{\prime}} & = & g {S^D} h (\gamma^{-1}) .
\end{array}
\right. \label{eq:twopolek}
\end{equation}

\section{Expectation values using the pole expansion}
\label{app:hampole}
In App.~\ref{app:expval} we derived expressions for the expectation values
of various operators in the Bloch states $| Q,k )$. We found that all
expectation values were expressed in terms of sums of matrix products. In
App.~\ref{app:pole} we showed that the asymptotic limit of a general sum
could be calculated. By doing a discrete Laplace transform of the sum and
analyzing the analytical structure of the transformed sum, we arrived at a
closed expression for the asymptotic behavior as a sum over just a few
matrices.

In this section we will combine the results of App.~\ref{app:expval} and
\ref{app:pole} and show how the particular sums in the expectation values of
App.~\ref{app:expval} can be analyzed with the technique of
App.~\ref{app:pole}. By doing this we will get rid of the unpleasant sums of
App.~\ref{app:expval} and replace them with simpler expressions describing
the asymptotic form of these expectation values in the limit where the
number of sites goes to infinity.

\subsection{The normalization}
We will begin with the simplest case, the norm as determined in 
Eq.~(\ref{eq:qknorm2})
\begin{equation}
( {{Q}^{\prime}}, k| Q,k ) = n { \, \mbox{tr} \left( \;   L_{{Q}^{\prime}} (
S_n^G - R_Q \hat{{\sl 1 }}^n )  \; \right) }, \label{eq:qknorm3} 
\end{equation}
with
\[
S_n^G = \Xi_n \left( \hat{{\sl 1 }}, R_Q, e^{ik} \hat{{\sl 1 }} \right) .
\]
In App.~\ref{app:pole} we arrived at two different expressions for the
asymptotic forms depending on if the momentum $k$ was zero or not. Let us
start with $k=0$. According to Eq.~(\ref{eq:twoasym0}), the sum $S_n^G$ then
has the asymptotic form
\[
S_n^G = B_G n + C_G .
\]
From Eq.~(\ref{eq:twopole0}) we directly get
\[
\left\{
\begin{array}{rcl}
B_G & = & g {R_Q^D} g \\
-(C_G -B_G) & = & {{g}^{\prime}} {R_Q^D} g + g {R_Q^D} {{g}^{\prime}} .
\end{array}
\right.
\]
The last term of Eq.~(\ref{eq:qknorm3}) is no sum and just gives an
additional matrix ${R_Q^D} g$ to the asymptotic form of $S_n^G$. Thus
\begin{equation}
( {{Q}^{\prime}}, k| Q,k ) = n { \, \mbox{tr} \left( \;   L_{{Q}^{\prime}} (
B_G n + C_G + {R_Q^D} g )  \; \right) } , \label{eq:gk} 
\end{equation}
where $k=0$ in this case.  Before going on to the case $k \neq 0$ we will
rewrite this formula on a more ``operator-like'' form. This can be done by
``pulling out'' the matrices $Q$ and ${{Q}^{\prime}}$ from the trace. We
note that $( {{Q}^{\prime}}, k| Q,k )$ in Eq.~(\ref{eq:gk}) has the form
\[
( {{Q}^{\prime}}, k| Q,k ) = \sum_\alpha { \, \mbox{tr} \left( \;
L_{{Q}^{\prime}} {M_\alpha} R_Q {N_\alpha}  \; \right) } , 
\]
with ${M_\alpha}$ and ${N_\alpha}$ square matrices on outer product form. By
doing a generalization of the tilde transformation of
Eq.~(\ref{eq:tildedef}) we can rewrite this as
\begin{eqnarray}
\sum_\alpha { \, \mbox{tr} \left( \;    L_{{Q}^{\prime}} {M_\alpha} R_Q
{N_\alpha}  \; \right) } & = & \sum_{\alpha} {{Q}^{\prime}}^* \widetilde{
{M_\alpha} {N_\alpha} } Q \nonumber \\  
& = & {{Q}^{\prime}}^* \left( \sum_\alpha \widetilde{ {M_\alpha} {N_\alpha}
} \right) Q \label{eq:gentilde} 
\end{eqnarray}
and we find that $\sum_\alpha \widetilde{ {M_\alpha} {N_\alpha} }$ gives a
closed expression for the norm operator, independent of $Q$ and
${{Q}^{\prime}}$ (but of course $k$-dependent). This transformation can be
accomplished by writing
\begin{eqnarray*}
\sum_{\alpha} { \, \mbox{tr} \left( \;   L_{{Q}^{\prime}}
{M_\alpha} R_Q {N_\alpha}  \; \right) } & = & ( {{Q}^{\prime}}^* \otimes 1
)^{(i,j),(k,l)} {M_\alpha}^{(k,l),(m,n)} ( 1 \otimes Q )^{(m,n),(o,p)} 
{N_\alpha}^{(o,p),(i,j)} \\
& = & {{Q}^{\prime}}^{{i,k}^*} \delta_{j,l} Q^{n,p} \delta_{m,o}
{M_\alpha}^{(k,l),(m,n)} {N_\alpha}^{(o,p),(i,j)} \\
& = & {{Q}^{\prime}}^{(i,k)^*} Q^{(n,p)} \left( {M_\alpha}^{T_{2341}}
\right) ^{(n,k),(j,m)} \left( {N_\alpha}^{T_{2341}} \right) ^{(j,m),(p,i)}
\\ 
& = & {{Q}^{\prime}}^{(i,k)^*} Q^{(n,p)} ( ( {M_\alpha}^{T_{2341}}
{N_\alpha}^{T_{2341}} )^{T_{3241}})^{(i,k),(n,p)} \\
& \equiv & {{Q}^{\prime}}^* \widetilde{{M_\alpha} {N_\alpha}} Q ,
\end{eqnarray*}
where the generalized transpose $M^{T_{i,j,k,l}}$ is defined in
Eq.~(\ref{eq:gentr}). We can thus define a ${{Q}^{\prime}}$ and $Q$
independent matrix $G(k,n)$ for $k=0$ by
\begin{displaymath}
({{Q}^{\prime}},k | Q,k) = {{Q}^{\prime}} G(k,n) Q ,
\end{displaymath}
where we determine $G(k,n)$ from Eq.~(\ref{eq:gk}) and the generalized tilde
transformation Eq.~(\ref{eq:gentilde}).

Likewise we can derive the expression for $G(k,n)$ for $k \neq 0$. This is
done in the same way by using the formulas Eq.~(\ref{eq:twopolek}) and
(\ref{eq:twoasymk}). The sum $S_n^G$ this time has the asymptotic form
\[
S_n^G = C_G + {{C}^{\prime}}_G ,
\]
where we have assumed $n$ such that $e^{ikn}=1$. We now find from 
Eq.~(\ref{eq:twopolek}) that
\begin{equation}
\left\{
\begin{array}{rcl}
C_G & = & g {R_Q^D} h ( e^{ik} ) \\
{{C}^{\prime}}_G & = & h ( e^{-ik} ) {R_Q^D} g .
\end{array}
\right.
\label{eq:cc0}
\end{equation}
The last term of Eq.~(\ref{eq:qknorm3}) is independent of $k$ and therefore
gives the same contribution, ${R_Q^D} g$, as before. From Eq.~(\ref{eq:cc0})
and the generalized tilde transformation we can calculate $G(k,n)$ also for
$k \neq 0$.

\subsection{The Hamiltonian}
Now we calculate the pole expansion of the Hamiltonian in
Eq.~(\ref{eq:qkham}). Let us start with $k \neq 0$ this time. We will
demonstrate the procedure for the term $H_A({{Q}^{\prime}},Q)$, just to
illustrate the three-partition case. For the rest of the terms we will, for
completeness, just list the results.

For $H_A$ we have from Eq.~(\ref{eq:qkham})
\[
H_A({{Q}^{\prime}},Q) = n { \, \mbox{tr} \left( \;   L_{{Q}^{\prime}}
S^A_{n-2}  \; \right) } , 
\]
where
\[
S^A_{n-2} = \Xi_{n-2} ( e^{-i k} \hat{{\sl 1 }}, R_Q, \hat{{\sl 1 }},
{\hat{S}^D \hat{S}^D}, \hat{{\sl 1 }} ) , 
\]
with the asymptotic form from Eq.~(\ref{eq:3asymk})
\[
S^A_{n-2} = B_A (n-2)  + C_A + \gamma^{-2} {{C_A}^{\prime}} ,
\]
where we have assumed $n$ such that $e^{i k n} = 1$. From
Eq.~(\ref{eq:3polek}) we get 
\begin{displaymath}
\left\{
\begin{array}{rcl}
B_A & = & h (e^{-ik}) {R_Q^D} g {\hat{S}^D \hat{S}^D} g  \\
-(C_A-B_A) & = & e^{-ik} {{ h }^{\prime}} (e^{-ik}) {R_Q^D} g {\hat{S}^D
\hat{S}^D} g + h (e^{-ik}) {R_Q^D} {{g}^{\prime}} {\hat{S}^D \hat{S}^D} g \\
& & \mbox{} + h (e^{-ik}) {R_Q^D} g {\hat{S}^D \hat{S}^D} {{g}^{\prime}} \\
{{C_A}^{\prime}} & = & g {R_Q^D} h (e^{ik}) {\hat{S}^D \hat{S}^D} h (e^{ik})
. 
\end{array}
\right.
\end{displaymath}
We now have
\begin{displaymath}
H_A({{Q}^{\prime}},Q) = n { \, \mbox{tr} \left( \;   L_{{Q}^{\prime}} \left(
B_A (n-2) + C_A + \gamma^{-2} {{C}^{\prime}}_A \right)  \; \right) }.  
\end{displaymath}
We transform this as we did with the norm using Eq.~(\ref{eq:gentilde}) to
get the $H_A$-operator 
\begin{displaymath}
H_A({{Q}^{\prime}},Q) = {{Q}^{\prime}} H_A(k,n) Q .
\end{displaymath}
Note the convention used here. We write the $m^2 \times m^2$ matrix operator
$H_A$, which is independent of $Q$ and ${{Q}^{\prime}}$ (but depends on $k$
and $n$), as $H_A(k,n)$ and the $({{Q}^{\prime}},Q)$-dependent expectation
value, which of course also depends on $k$ and $n$, as
$H_A({{Q}^{\prime}},Q)$ .

The matrix operator $H_B$ we get from $H_A$ by using Eq.~(\ref{eq:hastar})
\begin{eqnarray}
H_B({{Q}^{\prime}},Q) & = & \left( H_A(Q,{{Q}^{\prime}}) \right)^*
\nonumber\\ 
& = & \sum_\alpha Q^{(i,j)} \widetilde{ {M_\alpha} {N_\alpha}
}^{(i,j),(k,l)^*} {{Q}^{\prime}}^{(k,l)^*} \nonumber \\
& = & {{Q}^{\prime}}^* \left( \sum_\alpha \widetilde{ {M_\alpha} {N_\alpha}
}^\dagger \right) Q , \label{eq:hadagger} 
\end{eqnarray}
so that $H_B(k,n) = H_A^\dagger (k,n)$.

We now do the same procedure for the rest of the sums in
Eq.~(\ref{eq:qkham}) and then later also for the case $k=0$. The result for
the coefficients $A$, $B$, $C$ and ${{C}^{\prime}}$ in the asymptotic
expansion of different cases are listed below. Were a coefficient is not
present, it is zero.  Apart from the expression Eq.~(\ref{eq:hadagger}) we
can also get the asymptotic form of the sum in $H_B$, $S_{n-2}^B = \Xi_{n-2}
( \hat{{\sl 1 }}, {\hat{S}} {\hat{S}}, \hat{{\sl 1 }}, R_Q, e^{ik} \hat{{\sl
1 }} ) $ directly from the pole expansion as
\[
\left\{
\begin{array}{rcl}
B_B & = & g {R_Q^D} g {\hat{S}^D \hat{S}^D} h (e^{ik}) \\
-(C_B - B_B) & = & e^{ik} g {R_Q^D} g {\hat{S}^D \hat{S}^D} {{ h }^{\prime}}
(e^{ik}) + g {R_Q^D} {{g}^{\prime}} {\hat{S}^D \hat{S}^D} h (e^{ik}) \\ 
& & \mbox{} + {{g}^{\prime}} {R_Q^D} g {\hat{S}^D \hat{S}^D} h (e^{ik}) \\ 
{{C}^{\prime}}_B & = & h (e^{-ik}) {R_Q^D} h (e^{-ik}) {\hat{S}^D \hat{S}^D}
g . 
\end{array}
\right.
\]
The sum $S_{n-2}^F = \Xi_{n-2} ( \hat{{\sl 1 }}, {\hat{S}^D \hat{S}^D},
\hat{{\sl 1 }} )$ of $H_F$ gives 
\[
\left\{
\begin{array}{rcl}
B_F & = & g {\hat{S}^D \hat{S}^D} g \\
-(C_F - B_F) & = & {{g}^{\prime}} {\hat{S}^D \hat{S}^D} g + g {\hat{S}^D
\hat{S}^D} {{g}^{\prime}} . 
\end{array}
\right.
\]
The sum $S_{n-2}^C = \Xi_{n-2} ( e^{-ik} \hat{{\sl 1 }}, {\hat{S}} R_Q
{\hat{S}}, \hat{{\sl 1 }} )$ of $H_C$ yields
\[
\left\{
\begin{array}{rcl}
C_C & = & h (e^{-ik}) {{\hat{S}}^D R_Q^d {\hat{S}}^D} g \\
{{C}^{\prime}}_C & = & g {{\hat{S}}^D R_Q^d {\hat{S}}^D} h (e^{ik}) . 
\end{array}
\right.
\]
The sum in $H_D$, $S_{n-2}^D = \Xi_{n-2} ( e^{-ik} \hat{{\sl 1 }}, R_Q ,
\hat{{\sl 1 }} )$ can be derived from the relation $H_D({{Q}^{\prime}},Q) =
( H_C(Q,{{Q}^{\prime}}) )^*$ in the same way as we did for $H_B$:
\[
H_D(k,n) = H_C^\dagger(k,n) ,
\]
or directly from the pole expansion as
\[
\left\{
\begin{array}{rcl}
C_D & = & h (e^{-ik}) {R_Q^D} g \\
{{C}^{\prime}}_D & = & g {R_Q^D} h (e^{ik}) .
\end{array}
\right.
\]
Finally, for $H_E$ of Eq.~(\ref{eq:hefin}), which does not contain a sum, we
just replace the term $\hat{{\sl 1 }}^{n-2}$ by its asymptotic form, $g$,
and then perform the generalized tilde transform of Eq.~(\ref{eq:gentilde}).


The same procedure can be worked out for the Hamiltonian also when
$k=0$. However this time we will have to use the formulas in
Eq.~(\ref{eq:3pole0}) and (\ref{eq:twopole0}). The technique is analogous to
the $k \neq 0$ case and I will not list the results here. 

\subsection{The energy}

Collecting everything together we get for the whole Hamiltonian
\begin{displaymath}
H(k,n) = H_A(k,n) + H_B(k,n) - H_F(k,n) + H_C(k,n) + H_D(k,n) + H_E(k,n) 
\end{displaymath}
and for the energy
\begin{displaymath}
E_{{{Q}^{\prime}},Q} (k) = \frac{ ( {{Q}^{\prime}},k | H _{op}| Q,k ) }{
({{Q}^{\prime}},k | Q,k ) } = \frac{ {{Q}^{\prime}} H(k,n) Q }{
{{Q}^{\prime}} G(k,n) Q } 
\end{displaymath}
where $H(k,n)$ and $G(k,n)$ are square matrices. This is the result we
advertised in Eq.~(\ref{eq:qhkq}), (\ref{eq:qgkq}) and in
Eq.~(\ref{eq:hasy}) and (\ref{eq:gasy}).

Similar expressions for other expectation values like the total spin in
Eq.~(\ref{eq:qkstot}) can of course also be obtained.

\newpage 

\newpage

\newpage 

\newpage 

\narrowtext 

\begin{table}
\begin{tabular}{cdcc}
$\beta$			& $E_0$ 	& exact 	& best\ numerical\\ 
\hline  
$- \frac{1}{3} $	& $-$0.66666667	& $\frac{2}{3}$	& \\
$0$			& $-1$.40138	& -		&
$-1.401484038971(4)$ \\ 
$0.6$			& $-2$.9184	& -		& \\
$1.0$			& $-3$.98455	& $-4$		& \\ 
\end{tabular}
\caption{Ground state energy per site} 
\label{tab:gnd}
\end{table}

\end{document}